\documentclass[twocolumn,aps,pra,amsmath,amssymb,superscriptaddress,notitlepage,reprint,longbibliography,nofootinbib]{revtex4-1}

\usepackage{amsmath}

\input{KalmanPaper.rty}

\graphicspath{ {./} }
\begin{document}

\title{Optimal state estimation for cavity optomechanical systems}

\date{\today}

\author{Witlef Wieczorek}
\email{witlef.wieczorek@univie.ac.at}

\affiliation{University of Vienna, Faculty of Physics, Vienna Center for Quantum
  Science and Technology (VCQ), Boltzmanngasse 5, 1090 Vienna, Austria}

\author{Sebastian G. Hofer}
\email{sebastian.hofer@univie.ac.at}

\affiliation{University of Vienna, Faculty of Physics, Vienna Center for Quantum
  Science and Technology (VCQ), Boltzmanngasse 5, 1090 Vienna, Austria}

\affiliation{Leibniz University Hannover, Institute for Theoretical Physics, Institute
  for Gravitational Physics (Albert Einstein Institute), Callinstraße
  38, 30167 Hannover, Germany}

\author{Jason Hoelscher-Obermaier}
\email{jason.hoelscher-obermaier@univie.ac.at}

\affiliation{University of Vienna, Faculty of Physics, Vienna Center for Quantum
  Science and Technology (VCQ), Boltzmanngasse 5, 1090 Vienna, Austria}
\affiliation{Leibniz University Hannover, Institut f\"{u}r Gravitationsphysik, Institute for Gravitational Physics (Albert-Einstein-Institute), Callinstraße 38, 30167 Hannover, Germany}

\author{Ralf Riedinger}

\affiliation{University of Vienna, Faculty of Physics, Vienna Center for Quantum
  Science and Technology (VCQ), Boltzmanngasse 5, 1090 Vienna, Austria}

\author{Klemens Hammerer}

\affiliation{Leibniz University Hannover, Institute for Theoretical Physics, Institute
  for Gravitational Physics (Albert Einstein Institute), Callinstraße
  38, 30167 Hannover, Germany}

\author{Markus Aspelmeyer}

\affiliation{University of Vienna, Faculty of Physics, Vienna Center for Quantum
  Science and Technology (VCQ), Boltzmanngasse 5, 1090 Vienna, Austria}

\begin{abstract}
  We demonstrate optimal state estimation for a cavity optomechanical system through Kalman filtering. By taking into account nontrivial experimental noise sources, such as colored laser noise and spurious mechanical modes, we implement a realistic state-space model. This allows us to obtain the conditional system state, \ie{}, conditioned on previous measurements, with minimal least-square estimation error. We apply this method for estimating the mechanical state, as well as optomechanical correlations both in the weak and strong coupling regime. The application of the Kalman filter is an important next step for achieving real-time optimal (classical and quantum) control of cavity optomechanical systems.
\end{abstract}
\maketitle

\emph{Introduction}.---State estimation is a crucial task at the heart of control theory, both in the classical \cite{Stengel1994} and in the quantum domain \citep{Wiseman2010}. For Gaussian systems, real-time state estimation can be achieved in an optimal manner using Kalman-Bucy filtering \citep{kalman_new_1960,kalman_new_1961}. Since many physical systems are approximately Gaussian, Kalman filtering has been successfully implemented for a broad range of uses, for example for navigation and tracking in aeronautics (including the Apollo project and the Global Positioning System GPS) \citep{grewal_applications_2010}, as well as in the physical sciences, such as for suspension noise cancellation in gravitational wave detection \citep{Finn2001}, Heisenberg limited atomic magnetometry \citep{geremia_quantum_2003} or quantum-enhanced optical-phase tracking \citep{yonezawa_quantum-enhanced_2012,Tsang2009,Tsang2009a,Wheatley2010}. In this Letter we introduce a new domain of applications by implementing Kalman filtering for cavity optomechanical systems. These systems represent a versatile light-matter interface in which optomechanical interactions inside optical or microwave cavities allow control over optical and mechanical degrees of freedom. While the first investigations go back to the late 1960s in the context of gravitational wave detectors \citep{Braginsky1967a,braginsky_quantum_1995}, it is only the last few years that have seen the development of a completely new generation of micro- and nano-optomechanical solid-state devices with fast growing application areas from classical sensing to quantum information processing \citep{Aspelmeyer2014}.

State estimation of a cavity-optomechanical system in real time is key for optimal state control and verification. The outstanding challenge is to obtain reliable information on the mechanical subsystem. In optomechanics, this is done through an optical cavity field, which imposes both additional noise and dynamical back-action effects that have to be taken into account. Until now, reconstructions of the mechanical dynamics have focused either on statistical properties \citep{Paternostro2006,Palomaki2013} or, for real-time reconstructions, on regimes of sufficiently weak coupling and negligible dynamical back-action effects \citep{Rugar1991,Hadjar1999,Briant2003a,Iwasawa2013}. The information obtained from real-time estimation about the mechanical quadratures can be used for active feedback control of the mechanical resonator \citep{Briant2003a,Iwasawa2013,Wilson2014}. However, the validity of these reconstruction schemes breaks down when either coupling strength, dynamical back-action effects or noise become strong. Our Kalman filtering approach overcomes this limitation and allows us to demonstrate real-time optimal state estimation for cavity optomechanical systems operating in arbitrary parameter regimes. From a quantum physics perspective, the Kalman filter solves the stochastic Schrödinger equation---a stochastic, nonlinear generalization of the Schrödinger equation---which is the canonical way to describe quantum systems subject to a continuous measurement via coupling to electromagnetic fields \citep{carmichael_open_1993,gardiner_quantum_2004,bouten_introduction_2007,Wiseman2010}. These concepts and their application to mechanical systems have been the subject of extensive theoretical research \citep{mancini_optomechanical_1998,doherty_feedback_1999,hopkins_feedback_2003,danilishin_creation_2008,muller-ebhardt_quantum-state_2009}, but no experiments in the context of cavity-optomechanics have been conducted so far.

\emph{Kalman Filter}.---In quantum theory, just as in classical theories, a continuously observed system can be described by a \textit{conditional} \textit{state} \citep{Wiseman2010}, \ie{}, a state that incorporates the total amount of knowledge that an observer has extracted from her set of measurements. Discarding this knowledge yields the \textit{unconditional} \textit{state}, which is an incoherent mixture of all possible conditional states. Our goal is to find the (multipartite) conditional states of the full cavity-optomechanical system including mechanical and optical subsystems. We restrict ourselves to Gaussian dynamics and measurements, which is a valid assumption for the existing realizations of optomechanical systems \citep{Aspelmeyer2014}. For this case, it has been shown \citep{belavkin_optimal_1980} that the problem of finding the conditional state can be mapped to a classical estimation problem, which is solved by a Kalman filter. It produces a real-time state estimate from a continuous measurement trajectory, which is optimal in the sense of minimizing the mean square estimation error. We describe the system by the following (linear) state-space model
\begin{subequations}
  \label{eq:1}
  \begin{align}
    \ts{\dot{\vc x}} & =\At\ts{\vc x}+\ts{\vc w},\label{eq:2}\\
    \ts{\vc z} & =\Ct\ts{\vc x}+\ts{\vc v},\label{eq:3}
  \end{align}
\end{subequations}
where $\ts{\vc x}$ is a state vector in some appropriately chosen state space (\eg{}, the phase space of a harmonic oscillator), $\ts{\vc z}$ is the outcome of a linear measurement on the system, and $\ts{\vc w}$ and $\ts{\vc v}$ describe process and measurement noise, respectively. Both $\ts{\vc w}$ and $\ts{\vc v}$ are assumed to be zero-mean Gaussian white-noise processes, which obey $\Re\bigl(\ev{\ts{\vc w}\tss{\vc w}s^{\mathrm{T}}}\bigr)=W\delta(t-s)$ and $\Re\bigl(\ev{\ts{\vc v}\tss{\vc v}s^{\mathrm{T}}}\bigr)=V\delta(t-s)$, where $\delta$ is the Dirac $\delta$-function, $\ev{\cdot}$ denotes the expectation value with respect to the initial probability distribution describing system and noise, and $\Re(\cdot)$ the real part \footnote{Taking the real part of the covariance matrices that describe the noise processes is only necessary for quantum processes due to their non-commutative nature. Although it is not necessary for classical systems, we choose this explicitly real form for the sake of a consistent presentation.}. Process and measurement noise may be correlated, which is described by the cross-correlations $\Re\bigl(\ev{\ts{\vc w}\tss{\vc v}s^{\trans}}\bigr)=M\delta(t-s)$. $\At$ and $\Ct$ are real, matrix-valued functions that parametrize the state-space model and are fixed by the physical model of the system and measurement process. We aim to find the estimate $\ts{\hat{\vc x}}$ of $\ts{\vc x}$ that minimizes the mean square estimation error $\ev{\|\ts{\vc x}-\ts{\hat{\vc x}}\|^{2}}$ at a time $t$ when taking into account the measurement results $\{\tss{\vc z}s:0\leq s\leq t\}$. This yields $\ts{\hat{\vc x}}=\ev{\ts{\vc x}|\{\tss{\vc z}s:0\leq s\leq t\}}$, \ie{}, the conditional expectation value of $\ts{\vc x}$ given the full measurement record. Evaluating this for system \eqref{eq:1} yields the time-continuous version of the Kalman filter \citep{kalman_new_1961}
\begin{subequations} \label{eq:4}
  \begin{align}
    \ts{\dot{\hat{\vc x}}} & =\At\ts{\hat{\vc x}}+\ts K(\ts{\vc z}-\ts C\ts{\hat{\vc x}}),\label{eq:5}\\
    \ts{\dot{P}} & =\At\ts P+\ts P\At^{\mathrm{T}}+W-\ts KV\ts K^{\trans},\label{eq:6}
  \end{align}
\end{subequations}
where $\ts K=(\ts P\Ct^{\trans}+M)V^{-1}$ is the so-called Kalman gain and $\ts P=\Re\bigr(\ev{(\ts{\vc x}-\ts{\hat{\vc x}})(\ts{\vc x}-\ts{\hat{\vc x}})^{\trans}}\bigl)$ is the estimation-error covariance.

We can ascribe a quantum theoretical meaning to $\ts{\hat{\vc x}}$ and $\ts P$ by associating $\vc x$ with the Schrödinger operators that describe the quantum system, $\ts{\vc x}$ with the corresponding Heisenberg operators that evolve under \eqref{eq:2} (their Heisenberg--Langevin equations), and $\ts{\vc z}$ with an operator-valued output process \citep{belavkin_optimal_1980,bouten_introduction_2007}. If $\ts{\hat{\rho}}$ is the Gaussian state conditioned on a continuous measurement of $\ts{\vc z}$, we have $\ts{\hat{\vc x}}=\tr{\vc x\ts{\hat{\rho}}}$ and $\ts P=\Re\bigl(\tr{\vc x\vc x^{\trans}\ts{\hat{\rho}}}\bigr)-\ts{\hat{\vc x}}\ts{\hat{\vc x}}^{\trans}$, \ie{}, the \emph{symmetrized} covariance matrix of $\ts{\vc x}$ with respect to $\hat{\rho}$. In other words, the conditional Gaussian state $\ts{\hat{\rho}}$ is parametrized by $\ts{\vc{\hat{x}}}$ and $\ts P$ \footnote{One can also adopt a quantum-optical interpretation of $\ts{\hat{\vc x}}$. Formally integrating (and assuming vanishing initial conditions) gives $\ts{\hat{\vc x}}=\int_{-\infty}^{t}\mathcal{K}(t,s)\tss{\vc z}s\dd s$ with an integral kernel $\mathcal{K}$ depending on $\ts A$, $\ts C$ and $\ts K$. Thus $\ts{\hat{\vc x}}$ is formally equivalent to an (unnormalized) bosonic mode extracted from the output process $\ts{\vc z}$. In a quantum-optical setting this could be for example a temporal light mode extracted from the output light of a cavity. }. By averaging over all possible trajectories of $\ts{\hat{\vc x}}$ we recover the unconditional state, whose covariance matrix $\Re\bigl(\ev{\ts{\vc x}\ts{\vc x}^{\trans}}\bigr)$ we can extract from the estimated data by noting that $\ev{(\ts{\vc x}-\ts{\hat{\vc x}})\ts{\hat{\vc x}}^{\trans}}=0$, and thus $\Re\bigl(\ev{\ts{\vc x}\ts{\vc x}^{\trans}}\bigr)=\ts P+\ev{\ts{\hat{\vc x}}\ts{\hat{\vc x}}^{\trans}}$ \citep{Wiseman2010,hofer_entanglement-enhanced_2015}.

\begin{figure}[tb]
  \centering \includegraphics[width=\columnwidth]{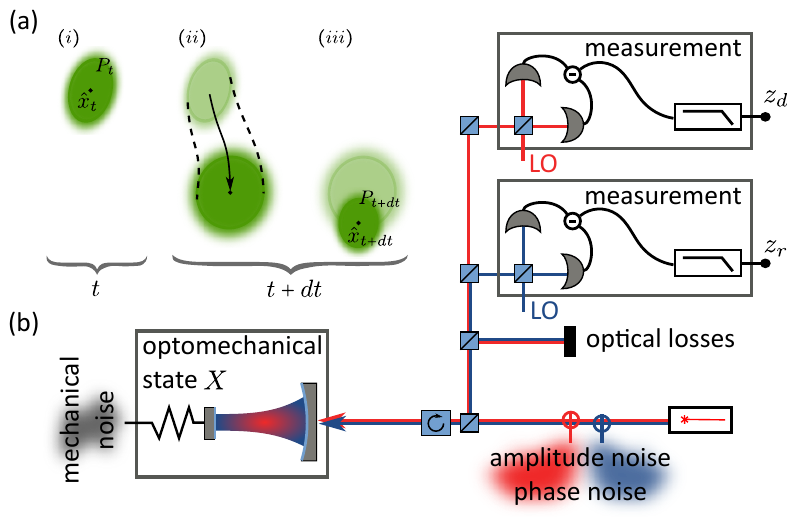} \caption{(color online) Kalman filter for cavity-optomechanical systems. (a) Working principle of the Kalman filter: (i) The conditional state is depicted by the green phase-space ellipse, which (ii) evolves in time according to the system dynamics. (iii) After a time $dt$ a Bayesian update is applied based on the measurement outcome to find the new conditional state. This procedure minimizes the mean-square estimation error, which makes the Kalman filter optimal for real-time state estimation. (b) Schematic of the experiment: The optomechanical cavity is driven by two laser beams each of which carry amplitude- and phase-noise. The mechanical motion is typically driven by Brownian noise. After their interaction with the cavity the optical fields are detected by two independent homodyne measurements (signals $z_{\protect\la}$ and $z_{\protect\lb}$), which themselves are subject to optical losses and noise. Building an accurate Kalman filter requires appropriate modeling of all relevant noise sources.}
  \label{fig:fig1}
\end{figure}

The Kalman filter equations \eqref{eq:4} describe how the conditional state is iteratively updated (\fref{fig:fig1}a). First, the estimate $\ts{\hat{{\vc x}}}$ and the covariance $\ts P$ are propagated for an infinitesimal time interval $dt$ {[}first term in \eqref{eq:5} and first three terms in \eqref{eq:6}{]} according to the state-space model \eqref{eq:2}. Second, the measurement outcome is incorporated as a Bayesian update that corrects the value of the estimate $\ts{\hat{{\vc x}}}$ and contracts the covariance ellipse {[}last terms in \eqref{eq:5} and \eqref{eq:6}{]}. The updated values are again propagated by $dt$ and the procedure is repeated.

\emph{The model.}---We consider a typical cavity-optomechanical architecture (Fig.~\ref{fig:fig1}b), in which a Fabry-P\'erot cavity (resonance frequency $\omega_{c}$), coupled to a single mechanical mode \footnote{The generalization to several mechanical modes is straight-forward, and, in fact, is included in the full model of our system.}, is driven by two laser fields (at frequencies $\omega_{0,\la}$, $\omega_{0,\lb}$). The ``resonant'' beam ($\omega_{\lb}=\omega_c$) acts as a weak probe of the cavity length to stabilize the laser frequency with respect to the cavity resonance; the ``detuned'' beam ($\omega_{\la}\neq\omega_c$) induces dynamical back-action effects, \eg{}, for laser cooling. This captures all relevant scenarios applied in typical optomechanics experiments. The mechanical element has a resonance frequency $\om$ and energy damping rate $\gamma_m$. Both cavity modes exhibit decay at a (half width at half maximum) rate $\kappa=\kappa_1+\kappa_2$, where $\kappa_1$ describes the input coupler and $\kappa_2$ accounts for spurious photon losses. The system is described by the (linearized) quantum Langevin equations \cite{giovannetti_phase-noise_2001,rabl_phase-noise_2009,abdi_effect_2011,ghobadi_optomechanical_2011}
\begin{subequations}\label{eq:7}
  \begin{align}
    \dot{q} & =\phantom{-}\om p\\
    \dot{p} & =-\om q-\gamma_mp+\sum_{\mathclap{i={r,d}}}g_i(\cos\theta_i x_i-\sin\theta_iy_i)+\xi,\\
    \dot{x}_i & =-\kappa x_i+\Delta_iy_i+g_i\sin\theta_iq+
                \sqrt{2\kappa_1}x_{i,1}^{\mathrm{in}}\nonumber\\
            &\hspace{3.7em}+\sqrt{2\kappa_2}x_{i,2}^{\mathrm{in}}+2\sqrt{\kappa_1} \, \delta\beta_i+|\alpha_{0,i}|\sin\theta_i\dot{\phi}_i,\\
    \dot{y}_{i} & =-\kappa y_{i}-\Delta_{i}x_{i}+g_{i}\cos\theta_iq+
              \sqrt{2\kappa_1}y_{i,1}^{\mathrm{in}}\nonumber \\
            & \hspace{8,7em}+\sqrt{2\kappa_2}y_{i,2}^{\mathrm{in}}+|\alpha_{0,i}|\cos\theta_i\dot{\phi}_i,
  \end{align}
\end{subequations}
where $q$, $p$ ($[q,p]=i$) describe position and momentum of the mirror, and $x_k$, $y_k$ with $[x_l,y_k]=i\delta_{lk}$ for $l,k\in\{\lb,\la\}$ respectively denote the amplitude and the phase quadrature of the cavity modes of the resonant and detuned beam. The optomechanical coupling to cavity mode $i$ is given by $g_{i}=\sqrt{2}g_{0}|\alpha_{0,i}|$ with $\alpha_{0,i}=\sqrt{2\kappa_1P_i/\hbar\omega_{0,i}}/(\kappa+\ii\Delta_{i})$, where $g_{0}$ is the single-photon coupling strength, ${P}_{i}$ is the corresponding driving laser power, and $\Delta_i=\omega_{0,i}-\omega_c$ is the detuning of the respective driving laser (at $\omega_{0,i}$) with respect to the cavity resonance frequency ($\omega_c$). The coupling of the mechanics to a thermal bath is modeled by a self-adjoint noise term $\xi$ with $\mean{\xi(t)\xi(s)+\xi(s)\xi(t)}=2\gamma_m(2\bar{n}+1)\delta(t-s)$ and $\bar{{n}}\approx k_{B}T/\hbar\om$ (the mean occupation number of the bath at temperature $T$). Optical shot noise is denoted by $x_{i}^{\mathrm{in}}$, $y_{i}^{\mathrm{in}}$ with variances $\mean{x_{i}^{\mathrm{in}}(t)x_{j}^{\mathrm{in}}(s)}=\mean{y_{i}^{\mathrm{in}}(t)y_{j}^{\mathrm{in}}(s)}=\tfrac{1}{2}\delta_{ij}\delta(t-s)$. Terms proportional to $\delta\beta_{i}$ and $\dot{\phi_{i}}$ describe classical amplitude and phase noise of the driving lasers \citep{rabl_phase-noise_2009,abdi_effect_2011,ghobadi_optomechanical_2011}.

Homodyne detection is used to independently measure the generalized quadratures $z_{i}$ of the reflected optical modes (\fref{fig:fig1}b). The cavity input-output relations yield
\begin{multline}
  z_{i}'=\left(\sqrt{2\kappa_1}\,x_{i}+x_{i}^{\mathrm{in}}+\delta\beta_{i}\right)\cos\varphi_{i}\\
  +\left(\sqrt{2\kappa_1}\,y_{i}+y_{i}^{\mathrm{in}}\right)\sin\varphi_{i},\label{eq:8}
\end{multline}
where $\delta\beta_{i}$ describes classical amplitude noise. We model optical losses and inefficient detection as beam-splitter losses parametrized by $\eta$. The measured quantities are rescaled to $z_{i}=\sqrt{1-\eta}\,z_{i}'+\sqrt{\eta}\,z_{i}^{\mathrm{in}}$, where $z_{i}^{\mathrm{in}}$ describes additional quantum noise independent of $x_{i}^{\mathrm{in}}$ and $y_{i}^{\mathrm{in}}$, \ie{}, $\mean{x_{i}^{\mathrm{in}}(t)z_{j}^{\mathrm{in}}(s)}=\mean{y_{i}^{\mathrm{in}}(t)z_{j}^{\mathrm{in}}(s)}=0$. Defining the vectors $\ts{\vc x}=(q(t),p(t),x_{\la}(t),y_{\la}(t),x_{\lb}(t),y_{\lb}(t))^{\trans}$ and $\ts{\vc z}=(z_{\la}(t),z_{\lb}(t))$ , equations \eqref{eq:7} and \eqref{eq:8} can be rewritten in the compact form \eqref{eq:1}.

Contrary to the idealizing assumptions made above, many of the noise sources in an actual experiment are frequency-dependent, here the laser amplitude- and phase-noise. This needs to be taken into account by properly extending the state space model. We incorporate three types of laser noise: (i) broadband laser noise originating from the laser itself, (ii) narrow-band Pound-Drever-Hall phase modulation in the resonant beam required for locking the laser to the cavity frequency, and (iii) narrow-band laser noise originating from the feedback loop of the laser lock. Each of these noise sources is experimentally characterized and is modeled independently to match the overall spectral characteristics (see \aref{sec:complete-state-space}). Furthermore, we extend the state space model to incorporate higher-order mechanical modes.

\begin{figure}[tb]
  \includegraphics[width=\columnwidth]{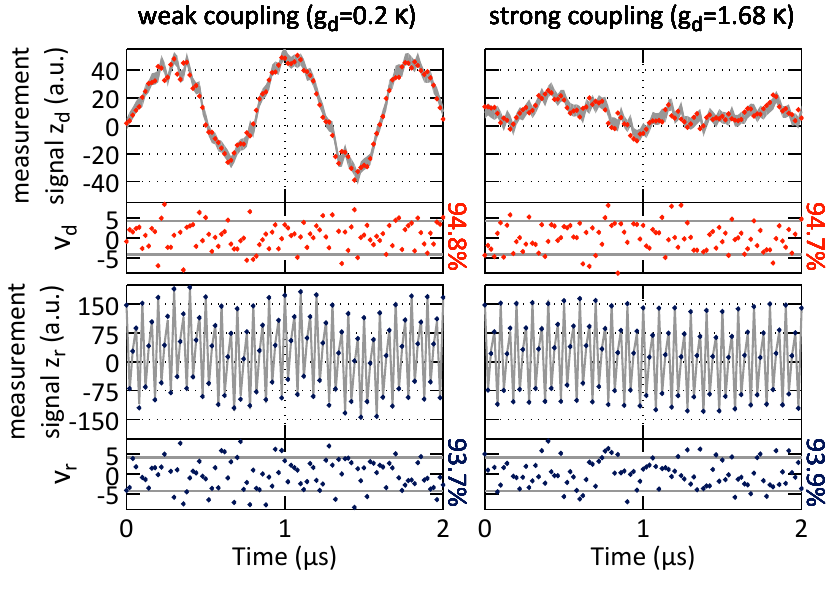} \caption{(color online) Measurement signals and Kalman-filter predictions. Shown are the measurement signals of the two homodyne detections of the detuned and resonant beam, $z_{\protect\la}(t)$ (red dots) and $z_{\protect\lb}(t)$ (blue dots), respectively, and their Kalman-filter predictions $\hat{z}_{i}(t)$ (gray line) both for the weak (left) and the strong coupling regime (right). Error bars of the prediction ($\pm2\sigma$) are indicated by the width of the gray line. Kalman filter innovations $\nu_{i}(t)=z_{i}(t)-\hat{z}_{i}(t)$ are plotted below each data set and demonstrate the accuracy of the implemented filter. To assess the performance of the filter, we calculate the fraction of normalized innovations that are contained in a two-sided $95\%$ confidence region ($\pm2\sigma$, indicated by the lines) of a zero-mean Gaussian distribution (besides each plot). The experiment was performed at room temperature with a micromechanical oscillator of $\omega_{m}=2\pi\times1.278\,$MHz, $\gamma_{m}=2\pi\times265\,$Hz, and optomechanical parameters $\kappa=0.34\,\omega_{m}$, $g_{0}=2\pi\times7.7\,$Hz (for details see \aref{sec:experiment}). For both coupling strengths of the detuned beam ($\Delta_{\protect\la}=\omega_{m}$) we use $\varphi_{\protect\la}\approx0$, $\Delta_{\protect\lb}=0$, $g_{\protect\lb}=0.2\,\kappa$ and $\varphi_{\protect\lb}=\pi/2$. Note that the fast oscillation of $z_{\protect\lb}(t)$, which is due to the 20 MHz Pound-Drever-Hall phase modulation for frequency locking, is taken into account by the Kalman filter. }
  \label{fig:fig2-1}
\end{figure}

\emph{Measurements and innovations.}---We use the recorded homodyne signals $z_{i}$ as input to the Kalman filter for estimation of the optomechanical state, which is done offline. \fref{fig:fig2-1} shows a $2\mu$s trace of the detector signals (corresponding to 100 sample points), along with the optimal measurement prediction. The prediction shows excellent qualitative agreement with the measured data both in the weak ($g_{\la}<\kappa$) and in the strong coupling regime ($g_{\la}>\kappa$). Quantitatively, the validity of the estimation is assessed by the innovation sequence $\ts{\boldsymbol{\nu}}=\ts{\vc z}-\ts C\ts{\vc{\hat{x}}}$, \ie{}, the difference between the predicted measurement $\vc{\hat{z}}_{t}=\ts C\ts{\vc{\hat{x}}}$ of the Kalman filter and the actual measurement outcome $\ts{\vc z}$. For an optimally working filter, $\ts{\boldsymbol{\nu}}$ must be a Gaussian zero-mean white noise process with a variance given by $\ev{\ts{\boldsymbol{\nu}}\ts{\boldsymbol{\nu}}^{\trans}}=\ts C\ts P\ts C^{\trans}+V$. We use this fact to fine-tune model parameters starting from their independently determined values. The statistics of $\ts{\boldsymbol{\nu}}$ of the resulting Kalman filter closely matches these criteria, hence demonstrating the accuracy of the filter (\fref{fig:fig2-1}; see also \aref{sec:kalm-filt-cons} for further statistical analysis).

\begin{figure}[tb]
  \center \includegraphics[width=\columnwidth]{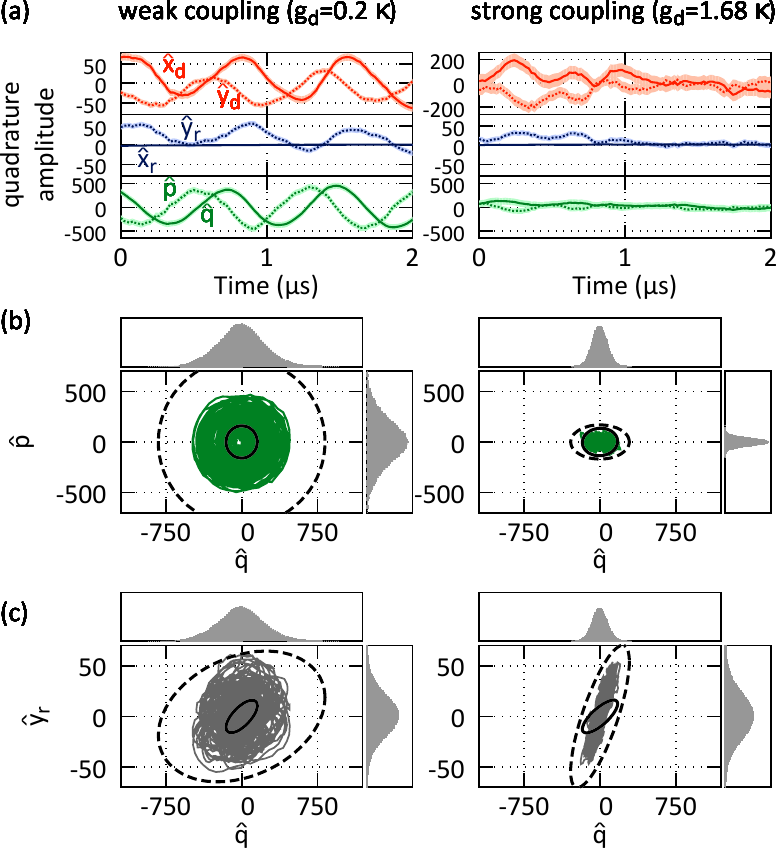} \caption{\label{fig:fig3-1}(color online) Estimating optomechanical quadratures. (a) Shown are Kalman-filter real-time estimates for the optical amplitude ($\hat{x}_i$, straight line) and phase ($\hat{y}_i$, dashed line) quadrature of the detuned (red, top) and resonant (blue, middle) beam along with the mechanical position ($\hat{q}$, straight green line) and momentum quadrature ($\hat{p}$, dashed green line) of the optomechanical system for the weak and strong coupling regime. The mechanical (b, green line) and optomechanical (c, gray line) phase space trajectories are estimated over a period of 100\,$\mu s$. A histogram along each quadrature is shown as side panel and estimated over $10$\,ms. The uncertainty ellipse of the unconditional (conditional) state is shown as dashed (straight) line. Note that the length of the shown trajectory is not sufficient to adequately represent the state's statistics. All units are given in terms of quadrature zero-point fluctuations (zpf). For our experimental parameters $q_{\mathrm{zpf}}=2.73\times10^{-16}$m, $p_{\mathrm{zpf}}=3.87\times10^{-19}\mathrm{kg\,m/s}$.}
\end{figure}

\emph{Estimation of optomechanical quadratures}.---Kalman filtering provides direct, real-time access both to the optical intracavity quadratures and to the mechanical degree of freedom in a cavity-optomechanical system (\fref{fig:fig3-1}a). In the weak coupling regime, the thermally driven mechanical motion and its coupling to the optical intracavity fields is visible. Clearly, the mechanical motion modulates both quadratures $x_{\la}$, $y_{\la}$ of the detuned beam \cite{Aspelmeyer2014}, which couples to the mirror via the optomechanical beam-splitter interaction. The situation is different for the resonant beam, whose amplitude quadrature $x_{\lb}$ contains shot noise only, while its phase quadrature $y_{\lb}$ couples to the mechanical position. 

The phase space representation captures the essence of Kalman filtering. The estimated mechanical quadratures rotate in phase space (\fref{fig:fig3-1}b). Their probability distribution along each mechanical quadrature is shown as histogram besides each axis and demonstrates the Gaussian nature of the micromirror motion. We compare the uncertainty ellipse of the unconditional (dashed line) and conditional (solid line) mechanical state, \ie{}, the area in which we expect with $95\%$ probability to find the mechanical quadratures. For a purely thermal state the area of the unconditional ellipse is proportional to the thermal occupation number $\bar{n}$. In the weak coupling regime, the information provided by the measurement update leads to a clear reduction in the uncertainty (factor of 27 in effective temperature), which is the optimal one for the given coupling strength \footnote{Recall that the optimality of the Kalman filter ensures that the conditional state uncertainty is minimized for a given coupling strength. }. In the strong coupling regime, laser cooling has already significantly diminished the thermally induced uncertainty of the unconditional state. In addition, the cavity dynamics introduces a notable ellipticity in the phase-space distribution \citep{genes_ground-state_2008}. The conditional state uncertainty is similar to the weak coupling situation. This is because for technical reasons the signal power at the homodyne detectors was kept constant for both coupling strengths, which means that the stronger detuned optical drive beam does not provide more information on the system state.

\Fref{fig:fig3-1}c shows real-time estimates of the optomechanical correlations between mechanical position and the phase quadrature of the resonant beam. Analogous to the mechanical phase space, the conditional state uncertainties are strongly reduced, reflecting the real-time information gain on the optomechanical correlations.

\emph{Conclusion}.---We have successfully implemented Kalman filtering for optimal state estimation of cavity optomechanical systems. Its  accuracy crucially relies on an accurate state space model of the specific experiment. The applications of this method in the domain of optomechanics are manifold. For example, Kalman filtering enables mechanical feedback control in the quantum regime. While in this work we operate the filter offline, its real time application in the frequency range investigated here is feasible using current field programmable gate array hardware (see \aref{sec:real-time-feedback}). The optimality of the filter guarantees that the reduction in conditional state uncertainty corresponds to the maximal cooling one can achieve through active feedback at this specific coupling strength. As a consequence, ground state cooling is readily achievable by combining Kalman filtering with measurements in the strong cooperativity regime \citep{hofer_entanglement-enhanced_2015}. This regime has been reached in current experiments \citep{brennecke_2008,Murch_observation_2008,purdy_observation_2013,Wilson2014}. In our case, it requires cryogenic cooling of the mechanical environment to 300\,mK and quality factors above $10^6$. As another example, mechanical sensing requires precise knowledge of the system dynamics in the absence of the external impetus, which is equivalent to the task of implementing the optimal estimator, \ie{}, the Kalman filter. The same is true for the task of characterizing or reconstructing an optomechanical quantum state (for example in terms of entanglement), where the relevant information is often encoded in the covariance matrix $P_{t}$. One fascinating prospect there is the generation of entanglement of macroscopic test masses through measurement \cite{muller-ebhardt_entanglement_2008,muller-ebhardt_quantum-state_2009}. In summary, Kalman filtering adds a significant performance advantage for classical and quantum control of cavity optomechanical systems.

\begin{acknowledgments}
  We thank Gerald Matz and Martin Siegele for discussions and Simon Gr\"oblacher for support with microfabrication. We acknowledge support by the European Commission (SIQS, iQOEMS, ITN cQOM, ThermiQ), the European Research Council (ERC QOM), the Austrian Science Fund (FWF): project numbers {[}Y414{]} (START), {[}F40{]} (SFB FOQUS), the Vienna Science and Technology Fund (WWTF) under Project ICT12-049 and the Centre for Quantum Engineering and Space-Time Research (QUEST). W.W. acknowledges support by a Feodor Lynen fellowship of the Alexander von Humboldt-Foundation and a Marie-Curie IEF of the European Commission. S.\,G.\ H.,\ J.~H.\,O.\ and R.\ R.~are supported by the Austrian Science Fund (FWF): project number {[}W1210{]} (CoQuS). Computations have been conducted in parts on the Vienna Scientific Cluster (VSC).
\end{acknowledgments}

\appendix

\section{State space models}
\label{sec:state-space-models}

\subsection{Systems driven by white noise}

State space models and Kalman filters based on them are widely used in classical signal processing \cite{bar-shalom_estimation_2001,heijden_classification_2005,stengel_optimal_1994}. Here we apply them to an optomechanical experiment in a way that remains applicable for quantum experiments.

Consider a classical Gaussian system with state vector~$\ts{\vc x}$ that is continuously monitored by linear measurements with outcomes~$\ts{\vc z}$. In general, both the evolution of the state vector as well as the measurement are affected by stochastic noise called \emph{process noise~}$\ts{\vc w}$ and \emph{measurement noise~}$\ts{\vc v}$, respectively. Additionally, both the state evolution as well as the measurement may be subject to a deterministic input~$\vc u$ ($\eg$~a known force).
Then, the joint (stochastic) evolution of state vector~$\ts{\vc x}$ and measurement outcomes~$\ts{\vc z}$ are described by a \emph{state space model} of the form
\begin{subequations} \label{eq:SSM}
\begin{align}
\ts{\dot{\vc x}} & =\At\ts{\vc x}+\Bt\ts{\vc u}+\ts L\ts{\vc w}\label{eq:SSM_general_state}\\
\ts{\vc z} & =\Ct\ts{\vc x}+\Dt\ts{\vc u}+\ts{\vc v}.\label{eq:SSM_general_measurement}
\end{align}
\end{subequations}
Here, $\At,\Bt,\Ct,\Dt$ and $\ts L$ are (potentially time-dependent) matrices whose naming is given in Tab.~\ref{tab:SSsymbols}. To completely specify the state space model \eqref{eq:SSM}, we also have to specify the statistical properties of the noise. If process noise~$\ts{\vc w}$ and measurement noise~$\ts{\vc v}$ are white and Gaussian this amounts to specifying covariance matrices $W$ and $V$ for process and measurement noise, respectively, as well as their correlation matrix~$M$.

\subsection{Colored noise and linear filters}

\label{sec:model-color-noise}

The Kalman filter relies on a state space model of the form~\eqref{eq:SSM} for which the process and measurement noise are Gaussian and white. Our experiment, however, is subject to colored Gaussian noise, namely classical amplitude and phase noise of our laser. But we can extend the state space model of the actual optomechanical system to an equivalent larger state space model, which is only driven by white noise, as described in the following. The Kalman filter can then be applied to this extended state space model.

Consider an $n$-dimensional (colored) noise process $\ts{\xi}$ whose time evolution is described by $\ts{\dot{\xi}}=F\ts{\xi}+\ts{\zeta}$, where $\ts{\zeta}$ is white noise. Its spectrum is given by the rational function $S_{\xi}(\omega)=H(\omega)WH(-\omega)^{\trans}=p_{\xi}(\omega)/q_{\xi}(\omega)$, where $H(\omega)=-(\ii\omega+F)^{-1}$ is the transfer function of the process, $W$ is the covariance matrix of $\ts{\zeta}$, and $\mathrm{deg}(q_{\xi})=n$.
Consider now a system driven by the colored noise process $\ts{\vc{\xi}}$, \ie{}, of the form
\begin{equation}
\ts{\dot{x}}=\At\xt+\Bt\ts u+\ts w+\ts{\xi}.\label{eq:SSM_colored_noise_driven}
\end{equation}
To model this system by an equivalent state space model driven by white noise only, we can define $\ts y=(\ts x^{\trans},\ts{\xi}^{\trans})^{\trans}$ and extend the state-space model \eqref{eq:SSM_colored_noise_driven} to
\begin{equation}
\ts{\dot{y}}=\begin{pmatrix}\At & \boldsymbol{1}_{n}\\
0 & F
\end{pmatrix}\ts y+\left(\begin{array}{cc}
\Bt & 0\\
0 & \text{0}
\end{array}\right)\begin{pmatrix}\ts u\\
0
\end{pmatrix}+\begin{pmatrix}\ts w\\
\ts{\zeta}
\end{pmatrix}.\label{eq:SSM_extended_white_noise_driven}
\end{equation}
We use such a state space model extension to incorporate classical, non-white laser noise. For our experiment, we model noise of narrow Lorentzian line-shape as well as broadband colored laser noise.

In a similar way, we can model the effect of an electronic or digital filter of order $n$ on an input signal $\ts u$. In this case, $\ts{\zeta}$ is related to the input signal by $\ts{\zeta}=G\ts u$, and the output is given by $\ts y=H\ts{\xi}+D\ts u$ (where $F$, $G$, $H$ and $D$ are matrices appropriately chosen to describe the specific filter). Its transfer function, relating the output to the input signal, is given by $\mathcal{G}_{\xi}(s)=\ts y(s)/\ts u(s)=H\frac{1}{s\openone-F}G+D$ with $s=\ii\omega$.

\section{Experiment}
\label{sec:experiment}

\subsection{Setup}

\begin{figure*}
  \centering{}\includegraphics[width=0.9\textwidth]{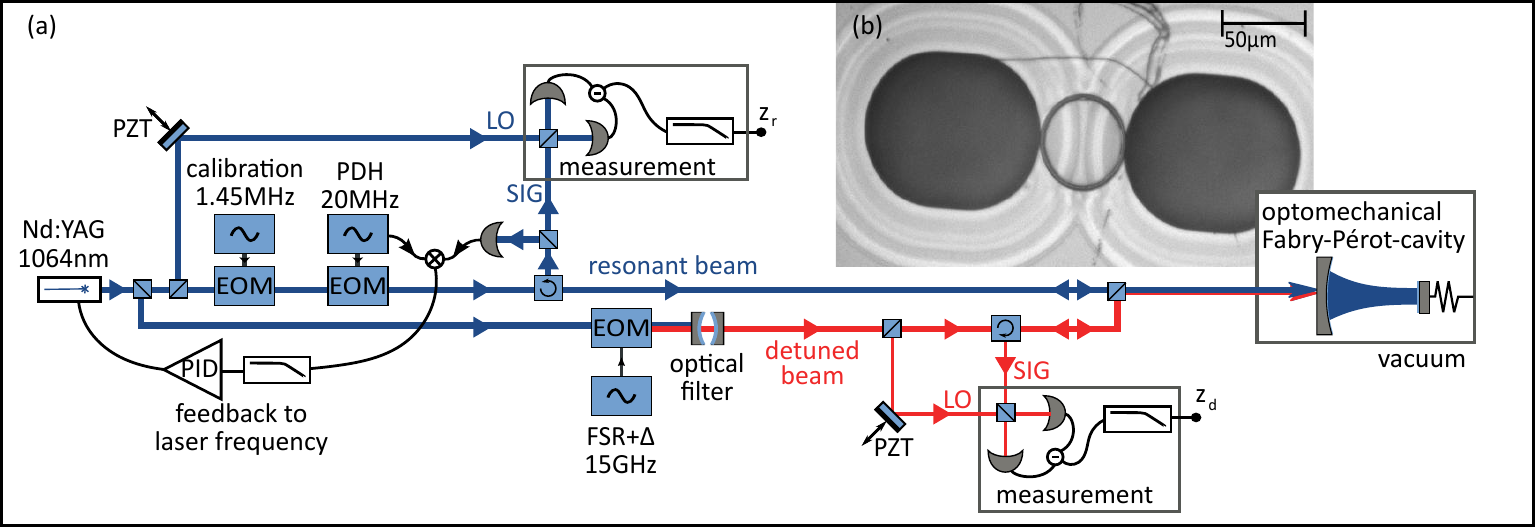} \protect\caption{\label{fig:experimental_setup}(a) Experimental setup. Laser light at $1064$\,nm is split into two beams. One of the beams (blue) is sent to two free space electro-optical modulators (EOMs), which add phase modulations at $1.45$\,MHz and $20$\,MHz. The modulation at $1.45$\,MHz is used for calibration of the optomechanical signal. The modulation at $20$\,MHz is used for locking the laser frequency to the optomechanical cavity (OMC) using a Pound-Drever-Hall (PDH) lock. This beam is therefore always resonant with the OMC and we refer
    to it as the ``resonant beam''.\protect \\
    ~~The second beam is sent to a fiber EOM which creates strong sidebands at approximately $15$\,GHz. One of the 1st order sidebands is used, whereas the carrier and all other sidebands are blocked by an optical filter consisting of a volume holographic grating and a broadband filter cavity. Hence, the resulting beam (red) is detuned with respect to the resonant beam (blue) by $\sim15$GHz which is approximately the free spectral range (FSR) of the OMC. This beam is therefore close to another longitudinal resonance of the OMC but can be detuned by
    an additional detuning $\Delta$ by changing the modulation frequency.\protect \\
    ~~The resonant and the detuned beam are combined on a polarizing beam splitter (PBS) into a single spatial mode but orthogonal polarizations. This mode is mode-matched to the OMC, which is located inside a vacuum environment. The returning beams (the signals) are separated from the incoming beams using optical circulators (half-wave plates, Faraday rotators and PBS) and sent to two homodyne detection setups. The relative phase of local oscillators (LO) and signals (SIG) can be stabilized using mirrors mounted on piezo-electric transducers~(PZT). The homodyne currents are low-pass filtered and the resulting signals $z_{\protect\la}$
    and $z_{\protect\lb}$ are digitized at $50$\,MHz.\protect \\
    (b) Optical micrograph of the mechanical oscillator. A doubly-clamped oscillator from SiN with a $\mathrm{Ta_{2}O_{5}/SiO_{2}}$Bragg mirror is used as the end mirror of the optomechanical cavity.}
\end{figure*}
The experimental setup is sketched in Fig.~\ref{fig:experimental_setup}. Our optomechanical cavity is a Fabry-P\'erot cavity with linewidth (half width at half maximum, HWHM) $\kappa\simeq440$\,kHz and free spectral range $\text{(FSR)}\simeq15$\,GHz. Its end mirror is a mechanical oscillator with frequency $\omega_{m}\simeq2\pi\cdot1.278\cdot10^{6}$\,Hz and linewidth (full width at half maximum, FWHM) $\gamma_m\simeq2\pi\cdot265$\,Hz. The oscillator consists of a doubly-clamped SiN~bridge with a $\mathrm{Ta_{2}O_{5}/SiO_{2}}$ Bragg mirror on top. The optomechanical cavity is placed inside vacuum ($p\simeq5\cdot10^{-6}$\,mbar); all measurements are taken at room temperature.

The mechanical oscillator interacts with two optical cavity modes in orthogonal polarizations and separated in frequency by one FSR. These two cavity modes are driven by two laser beams, one resonant and one red-detuned by approximately the mechanical frequency. Both beams are derived from a single Nd:YAG laser with a wavelength of 1064nm as described in more detail in Fig.~\ref{fig:experimental_setup}. The laser is locked to the optomechanical cavity using a Pound-Drever-Hall (PDH) lock in the resonant beam.

The reflected beams (the signal beams) are split off using wave plates and Faraday rotators and directed to two separate homodyne detection setups. The local oscillators for the homodyne detection are derived from the incoming beams. The path lengths of signal beams and local oscillators are carefully matched to ensure that signals and local oscillators always have the same instantaneous frequency such that the laser phase noise does not affect the homodyne detection. The relative phase of local oscillator and reflected (signal) beams can be locked using feedback to piezo-driven mirrors (PZT) in the local oscillator beam paths. This allows, for each returning signal beam, to measure an arbitrary generalized quadrature. For the detuned beam, the phase $\varphi_{\la}$ is scanned in time between $0$ and $\pi/2$, whereas for the resonant beam, the phase is fixed to $\varphi_{\lb}\simeq\pi/2$.

\subsection{Classical laser noise}

\subsubsection{Laser field}

We describe the extra-cavity laser fields of each beam by a displaced coherent state $\ket{\bt}$ with a fluctuating coherent amplitude $\beta(t)$ (note that we suppress the mode indices $\la$ and $\lb$ here and in the following). In a frame rotating at the laser frequency, we can write the coherent amplitude as
\begin{equation}
\bt=(\bo+\db(t))\ee^{-\ii\phi(t)},\label{eq:def-laser-fluctuations}
\end{equation}
where $\bt$ is a classical complex random variable with expectation value $\ev{\bt}=\bo\in\reals$. The amplitude and phase fluctuations $\db(t),\thinspace\phi(t)$ are assumed to be real, small ($\ev{\db^{2}}\ll\bo^{2},\ev{\phi^{2}}\ll1$), uncorrelated ($\ev{\db\thinspace\phi}=\ev{\db}\ev{\phi}$) and zero-mean ($\ev{\db}=\ev{\phi}=0$).
Here and in the following, we use 
\[
\ev{f(\beta)}\defeq\int_{\mathbb{C}}\dd^{2}\beta\,\,p(\beta)\thinspace f(\beta)
\]
to denote the classical expectation value of $f(\beta)$ and $p(\beta)$ for the probability density function of $\beta$ in the complex plane. Taking the classical fluctuations into account, we must therefore describe the extra-cavity laser fields by a mixed state
\[
\rho=\int\dd^{2}\beta\thinspace p(\beta)\ket{\beta}\bra{\beta}.
\]
The expectation value of an operator $O$ in state $\rho$ is then defined as
\[
\braket O\defeq\tr{\rho\,O}=\int\dd^{2}\beta\thinspace\,p(\beta)\bra{\beta}O\ket{\beta}.
\]
Assuming that we are only dealing with wide-sense stationary random processes, the noise power spectrum $S_{OO}(\omega)$ of an operator-valued noise process $O(t)$ is defined as \cite{clerk_introduction_2010}
\[
S_{OO}(\omega)=\int_{\reals}\dd t\ee^{\ii\omega t}\braket{O(t)O(0)}.
\]
To calculate noise power spectra of photocurrents we need to use the following properties of the annihilation and creation operators $b,b^{\dagger}$ of the extra-cavity field modes
\begin{eqnarray*}
b(t)\ket{\bt} & = & \bt\ket{\bt}\\
\bra{\bt}b(t)^{\dagger} & = & \bt^{*}\bra{\bt}\\
{}[b(t),b(t')^{\dagger}] & = & \delta(t-t'),
\end{eqnarray*}
where $\delta(t)$ denotes the Dirac delta distribution and the operators $b,b^{\dagger}$ have units of $\sqrt{\mathrm{Hz}}$.

\subsubsection{Amplitude noise}

\setcounter{paragraph}{0}

\paragraph{Measurement.}
We measure the amplitude noise of our laser by direct detection on an InGaAs photodiode with a detection bandwidth of up to 20\,MHz. The statistics of the detected photocurrent $I(t)$ is proportional to the statistics of the number operator $N=b^{\dagger}b$.

Using the definition \eqref{eq:def-laser-fluctuations} of the laser amplitude and phase fluctuations, we find for the noise power spectral density of the photon number in direct detection
\begin{equation}
S_{NN}(\omega)=\underset{\text{classical amplitude noise}}{4\bo^{2}\underbrace{S_{(\db)(\db)}(\omega)}}+\underset{\text{shot noise}}{(\underbrace{\bo^{2}+\ev{\db^{2}}}).}\label{eq:NPS-direct-detection}
\end{equation}
Note that we omitted DC-terms proportional to $\delta(\omega)$ in \eqref{eq:NPS-direct-detection}. Note also that the noise power spectral density of the shot noise is equal to the mean photocurrent.

Since all extra-cavity fields $\ket{\bt}$ are derived from the same laser field with state
\begin{equation}
\ket{\gt}=\ket{(\go+\dg(t))\ee^{-\ii\phi(t)}},
\end{equation}
we can write $\bt=r\gt$ such that $\bo=r\go$ and $\db(t)=r\thinspace\dg(t)$; here, $r^{2}=P/P_{\text{laser}}$ is the ratio of the power~$P$ of the field $\bt$ we measure to the power $P_{\text{laser}}$ of the laser. Therefore, we can write

\begin{eqnarray*}
S_{NN}(\omega) & \simeq & 4\bo^{2}S_{(\db)(\db)}(\omega)+\bo^{2}\\
 & \propto & \underset{\text{classical amplitude noise}}{\underbrace{4(P^{2}/P_{\text{laser}})S_{(\dg)(\dg)}(\omega)}}+\underset{\text{shot noise}}{\underbrace{P}}.
\end{eqnarray*}
For high optical powers $P$, shot noise is negligible and we can directly measure the classical amplitude noise, such that $S_{II}(\omega)\propto S_{NN}(\omega)\propto S_{(\db)(\db)}(\omega)$. By scaling appropriately with the ratio of the powers we can then calculate the classical amplitude noise power also for weaker fields.

\begin{figure}
  \centering{}\includegraphics{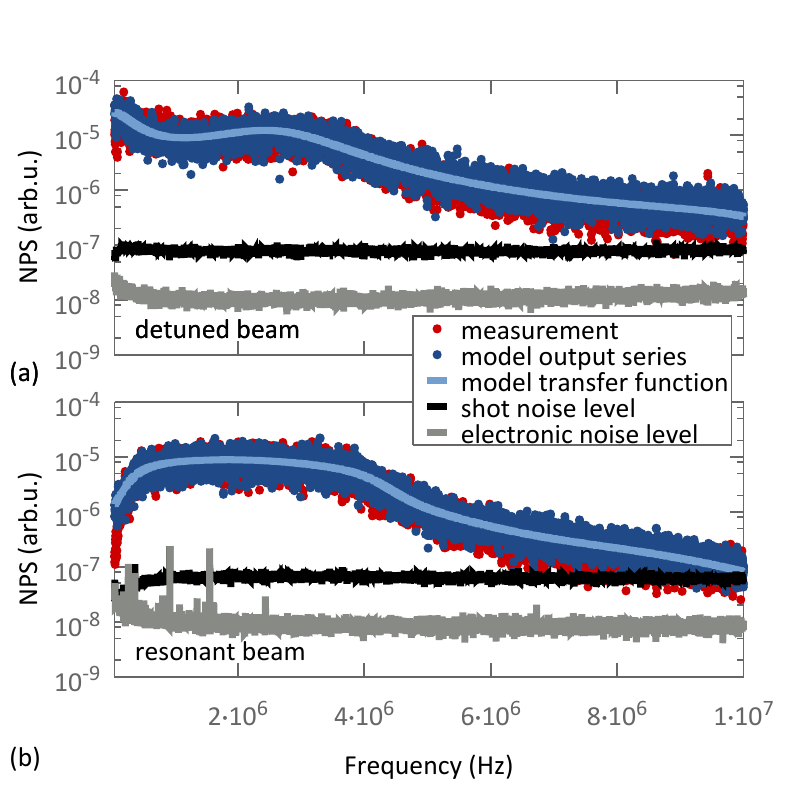} \protect\caption{\label{fig:colnoiseamp}Measured and modeled amplitude noise. The measurements (red points) show the photocurrent noise power spectral density $S_{II}(\omega)$ in direct detection for the detuned (a) and resonant (b) optical beam with respective shot noise (black line) and electronic noise levels (gray line). In the relevant frequency range, $S_{II}(\omega)$ is dominated by the contribution of the classical amplitude noise $S_{(\protect\db)(\protect\db)}(\omega)$. A state space model for the amplitude noise is identified based on these measurements. We plot the noise power spectrum of an exemplary output time series of these state space models (blue points) together with their transfer functions (straight lines).}
\end{figure}

\paragraph{Modeling.}%
Our goal is to find a state space model that reproduces the spectral dependence of the amplitude noise $S_{(\db)(\db)}(\omega)$ in a satisfactory way. Since, for high optical powers, the measured photocurrent $I(t)$ is directly proportional to the relevant amplitude noise signal $\db(t)$, we use $I(t)$ as input for identifying a suitable state space model. This is done using the MATLAB System Identification toolbox. The results are shown in \fref{fig:colnoiseamp}. 

\subsubsection{Homodyne detection with noise}

We use a standard homodyne setup in which local oscillator (mode 1) and signal (mode 2) are combined on a 50:50 beam splitter. The output modes 3 and 4 are detected and their photocurrents subtracted from each other. The statistics of the resulting difference current $I_{-}(t)$ is proportional to that of the photon number difference operator
\begin{eqnarray*}
N_{-} & \defeq & b_{3}^{\dagger}b_{3}-b_{4}^{\dagger}b_{4}=\ee^{-\ii\varphi}b_{1}^{\dagger}b_{2}+\ee^{\ii\varphi}b_{2}^{\dagger}b_{1}.
\end{eqnarray*}
Here, $\varphi$ is the additional (constant) phase that the local oscillator acquires with respect to the signal.

Now we assume the following states for the local oscillator (mode 1) and signal (mode 2):
\begin{eqnarray}
\ket{\psi_{1}} & = & \ket{(\bo+\db)\ee^{-\ii\phi}},\nonumber \\
\ket{\psi_{2}} & = & \ket{r(\bo+\db)\ee^{-\ii\phi}+\dx+\ii\dy}.\label{eq:homodyning-signal-field-state}
\end{eqnarray}
The signal is derived from the same laser as the local oscillator; therefore, its coherent amplitude has a contribution $r(\bo+\db)\ee^{-\ii\phi}$ with $r^{2}=P_{\text{sig}}/P_{\text{LO}}$, \ie{}, the ratio between the optical powers $P_{\text{sig}}$ and $P_{\text{LO}}$ of signal and local oscillator. The signal beam carries additional amplitude and phase quadrature fluctuations $\dx$ and $\dy$ due to the interaction with the optomechanical cavity, which are assumed to be zero-mean $\ev{\dx}=\ev{\dy}=0$. Note that, by writing the signal field state as a displaced coherent state \eqref{eq:homodyning-signal-field-state}, we neglect the possibility that the optomechanical interaction changes the quantum statistics of the output light field, e.g.,~leading to squeezing \cite{brooks_non-classical_2012,safavi-naeini_squeezed_2013,purdy_strong_2013}.
We then find (in dependence of the homodyning angle~$\varphi$)
\begin{eqnarray}
S_{N_{-}N_{-}}(\omega) & = & \underset{\text{shot noise}}{\underbrace{(\bo^{2}+\ev{\db^{2}})(1+r^{2})+\ev{\dx^{2}}+\ev{\dy^{2}}}}\nonumber \\
 &  & \underset{\text{signal noise}}{\underbrace{+4\bo^{2}\thinspace S_{x(\varphi)x(\varphi)}(\omega)}}\label{eq:NPS-homodyning}\\
 &  & +(r\thinspace\cos\varphi)^{2}\underset{\text{classical amplitude noise}}{\underbrace{16\bo^{2}\thinspace S_{\db\db}(\omega)}}\nonumber 
\end{eqnarray}
In \eqref{eq:NPS-homodyning} we again omitted DC-terms and terms that are of more than second order in the fluctuations ($\db,\phi,\dx,\dy)$.

Depending on the homodyning angle $\varphi$, we detect the noise $S_{x(\varphi)x(\varphi)}(\omega)$ of the signal in the generalized quadrature $x(\varphi)\defeq\cos(\varphi)\dx+\sin(\varphi)\dy$. For $\varphi=0$, in particular, we detect the amplitude fluctuations $S_{\dx\dx}(\omega)$ of the signal together with the common amplitude fluctuations of signal and local oscillator $S_{\db\db}(\omega)$. For $\varphi=\pi/2$, on the other hand, we detect the phase fluctuations $S_{\dy\dy}(\omega)$. The common phase fluctuations $\phi(t)$ of signal and local oscillator cancel as we aligned the beam paths to equal lengths such that both beams have the same instantaneous optical frequency at the detectors. For any value of the homodyning angle $\varphi$, we detect a constant shot noise background proportional to the combined optical power of signal and local oscillator field.

\subsubsection{Frequency and phase noise}

\setcounter{paragraph}{0}

\paragraph{Measurement.}

To measure the phase noise of our laser we use delayed self-homodyning
\cite{safavi-naeini_laser_2013}. The laser beam is split into two
beams (beams 1 and 2) of equal power. One of the resulting beams (beam
2) is then delayed by a time $\Delta T$ before both beams are recombined
on a 50:50 beam splitter. Both outputs of the 50:50 beam splitter
are detected and their photocurrents subtracted. At low frequencies,
where the large-amplitude phase fluctuations occur, the phase between
both beams is stabilized to $\varphi=\pi/2$.

The difference current $I_{-}$ is then proportional to the number
difference operator 
\[
N_{-}=b_{3}^{\dagger}b_{3}-b_{4}^{\dagger}b_{4}=\ii(b_{2}^{^{\dagger}}b_{1}-b_{1}^{\dagger}b_{2}),
\]
and we assume beam 1 and 2 to be in the states $\ket{\psi_{1}}=\ket{(\bo+\db(t))\ee^{-\ii\phi(t)}}$
and $\ket{\psi_{2}}=\ket{(\bo+\db(t+\Delta T))\ee^{-\ii\phi(t+\Delta T)}}$,
respectively. 

Assuming small phase fluctuations $\ee^{-\ii\phi(t)}\simeq1-\ii\phi(t)$
and neglecting terms of more than second order in the fluctuations,
we find the noise power spectrum of the difference current to be proportional
to

\begin{equation}
S_{N_{-}N_{-}}(\omega)=4\bo^{4}\thinspace S_{\Delta\phi\Delta\phi}(\omega)+2(\bo^{2}+\ev{\db^{2}}).\label{eq:NPS-accumulated-phase-noise}
\end{equation}
In \eqref{eq:NPS-accumulated-phase-noise}, $S_{\Delta\phi\Delta\phi}(\omega)$
is the noise power spectral density of the accumulated phase difference
$\Delta\phi(t)=\int_{t}^{t+\Delta T}d\tau\dot{\phi}(\tau)$ with $\dot{\phi}(\tau)$
the instantaneous optical frequency. This can be related \cite{riedinger_2013}
to the noise power spectrum of the fluctuating phase $\phi$ via
\[
S_{\Delta\phi\Delta\phi}(\omega)=4\sin^{2}\left(\frac{\omega\,\Delta T}{2}\right)S_{\phi\phi}(\omega),
\]
such that we get 

\begin{eqnarray*}
S_{N_{-}N_{-}}(\omega) & = & \underset{\text{classical phase noise contribution}}{\underbrace{16\bo^{4}\thinspace\sin^{2}\left(\frac{\omega\,\Delta T}{2}\right)\thinspace S_{\phi\phi}(\omega)}}\\
 &  & +\underset{\text{shot noise}}{\underbrace{2(\bo^{2}+\ev{\db^{2}})}}.
\end{eqnarray*}
The noise power spectrum of the phase noise $S_{\phi\phi}(\omega)$
can therefore be obtained from the spectrum of the detected difference
current by subtracting the shotnoise and dividing by the function
$\sin^{2}(\frac{\omega\,\Delta T}{2})$ which acts like a frequency-dependent
gain for the phase fluctuations. Note that for frequencies $\omega$
around integer multiples of $2\pi/\Delta T$, the interferometer is
insensitive to phase noise. To get an accurate phase measurement also
at these frequencies, the measurement has to be repeated with a different
delay $\Delta T$.

\paragraph{Modeling.}

\begin{figure}
\begin{centering}
\includegraphics{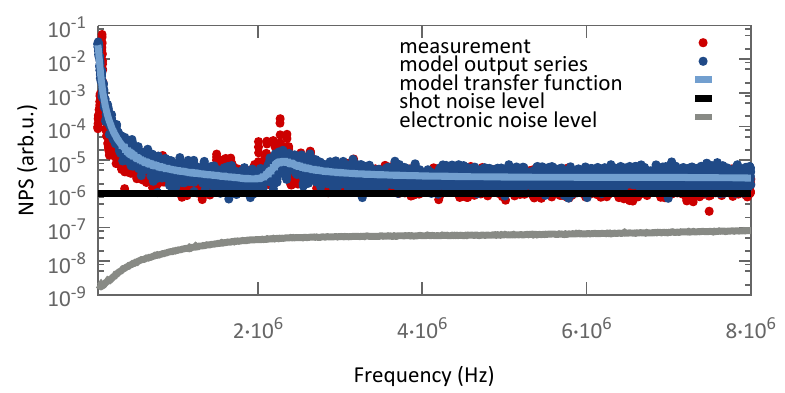}
\protect\caption{\label{fig:colnoisephase} \textbf{\uline{}}Measured and modeled frequency noise. The measurements (red points) show the photocurrent noise power spectral density $S_{I_{-}I_{-}}(\omega)$ obtained using delayed self-homodyning for the resonant optical beam. Also shown is the electronic noise level (gray line). In the relevant frequency range, $S_{I_{-}I_{-}}(\omega)$ is proportional to the classical frequency noise $S_{\dot{\phi}\dot{\phi}}(\omega)$. A state space model for the frequency noise is identified based on these measurements. We plot the noise power spectrum of an exemplary output time series of the state space model (blue points) together with its transfer function (straight lines). Note that the sharp resonance around $2.5\,$MHz are resonances of the piezoelectric transducer in the laser which we do not model separately.}
\par\end{centering}
\end{figure}

We actually need the fluctuating frequency $\dot{\phi}$ rather than the fluctuating phase $\phi$ as input to the optomechanical state space model. To obtain a state space model for the frequency noise $S_{\dot{\phi}\dot{\phi}}(\omega)$ from the photocurrent $I_{-}$ measured in delayed self-homodyning, we note that $S_{\dot{\phi}\dot{\phi}}(\omega)=\omega^{2}S_{\phi\phi}(\omega)$. Furthermore, we measure at optical powers which are high enough for the shot noise to be neglected such that we get
\begin{equation}
S_{II}(\omega)\propto\frac{\sin^{2}\left(\frac{\omega\,\Delta T}{2}\right)}{\omega^{2}}\thinspace S_{\dot{\phi}\dot{\phi}}(\omega).\label{eq:NPS-frequency-noise}
\end{equation}
We choose a small $\Delta T=27$\,ns such that we can linearize the $\sin^{2}$-term in \eqref{eq:NPS-frequency-noise} for the relevant frequencies below 5\,MHz and get $S_{I_{-}I_{-}}(\omega)\propto S_{\dot{\phi}\dot{\phi}}(\omega).$ Hence, for high optical powers and low frequencies the measured photocurrent $I_{-}$ is directly proportional to the frequency noise $\dot{\phi}$ and can be used as input for identifying a state space model in analogy to the amplitude noise (see \fref{fig:colnoisephase}).

\subsection{Spurious mechanical modes}

\begin{figure}
  \centering{}\includegraphics[width=1\columnwidth]{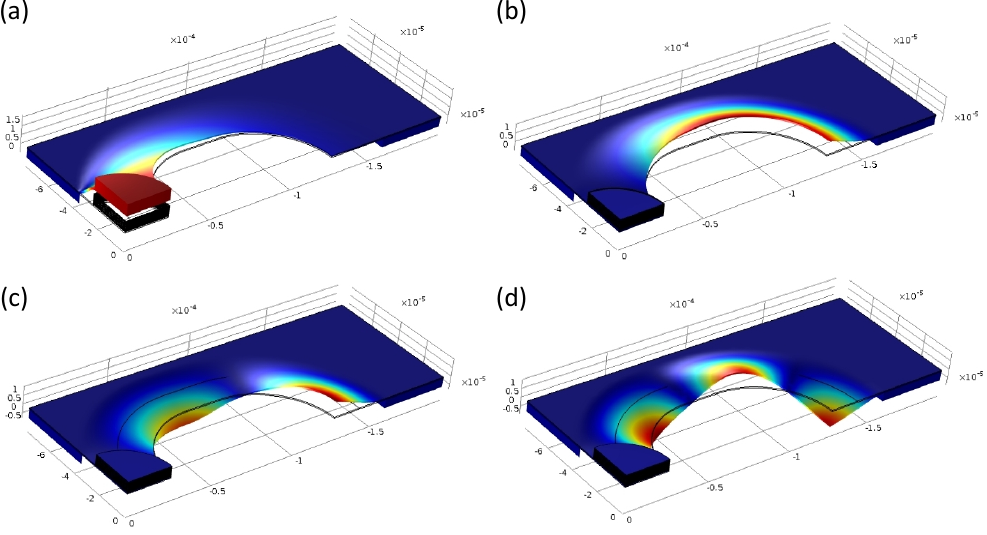} \protect\caption{\label{fig:FEM}Finite-element modelling of the utilized doubly-clamped mechanical oscillator. Shown is the displacement amplitude of (a) the fundamental mode with a modelled frequency of 1.28\,MHz and higher order modes with frequencies of (b) 2.31\,MHz, (c) 3.08\,MHz, (d) 4.31\,MHz.}
\end{figure}

In our experiment, we use a doubly-clamped mechanical oscillator [see \fref{fig:experimental_setup}(b)], which is coupled to two optical fields. This oscillator supports multiple mechanical modes, as is common with clamped oscillators. Some mechanical modes of our oscillator are shown in \fref{fig:FEM} which have been calculated by finite element modelling.

\begin{figure}
  \centering{}\includegraphics{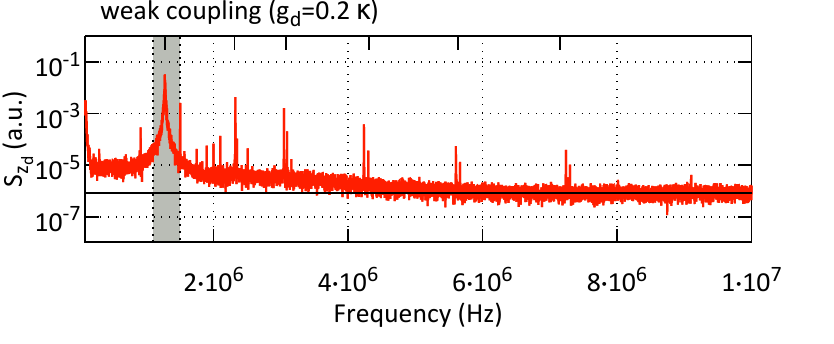} \protect\caption{\label{fig:NPSHD}Noise power spectrum of signal $z_\la$ in the weak coupling regime. The fundamental mechanical mode at a frequency of 1.278\,MHz is marked with the vertically gray-shaded region. Other peaks at frequencies of 2.325\,MHz, 3.05\,MHz, 4.237\,MHz, 5.604\,MHz and 7.24\,MHz can be attributed to mechancial modes calculated via finite element modelling and are marked at the top horizontal axis, cf.~\fref{fig:FEM}.}
\end{figure}

We are interested in coupling to the fundamental out-of-plane mechanical mode [\fref{fig:FEM}(a)], as it exhibits the largest optomechanical coupling strength. However, all other mechanical modes will also couple to the optical fields, albeit with lower coupling strength. \Fref{fig:NPSHD} shows the noise power spectrum of the homodyne signal $z_\la$ of the detuned beam, where we can attribute some peaks in the spectrum to mechanical modes. We also incorporate these other mechanical modes in the state space model of our experiment. Only then the Kalman filter will yield an accurate estimate of the state of the fundamental mechanical mode. The consistency of the state estimation can be checked by inspecting the innovation sequence, for details see \sref{sec:kalm-filt-cons}.

\section{Complete State Space Model}
\label{sec:complete-state-space}

\begin{figure*}
\begin{centering}
	\includegraphics[width=0.9\textwidth]{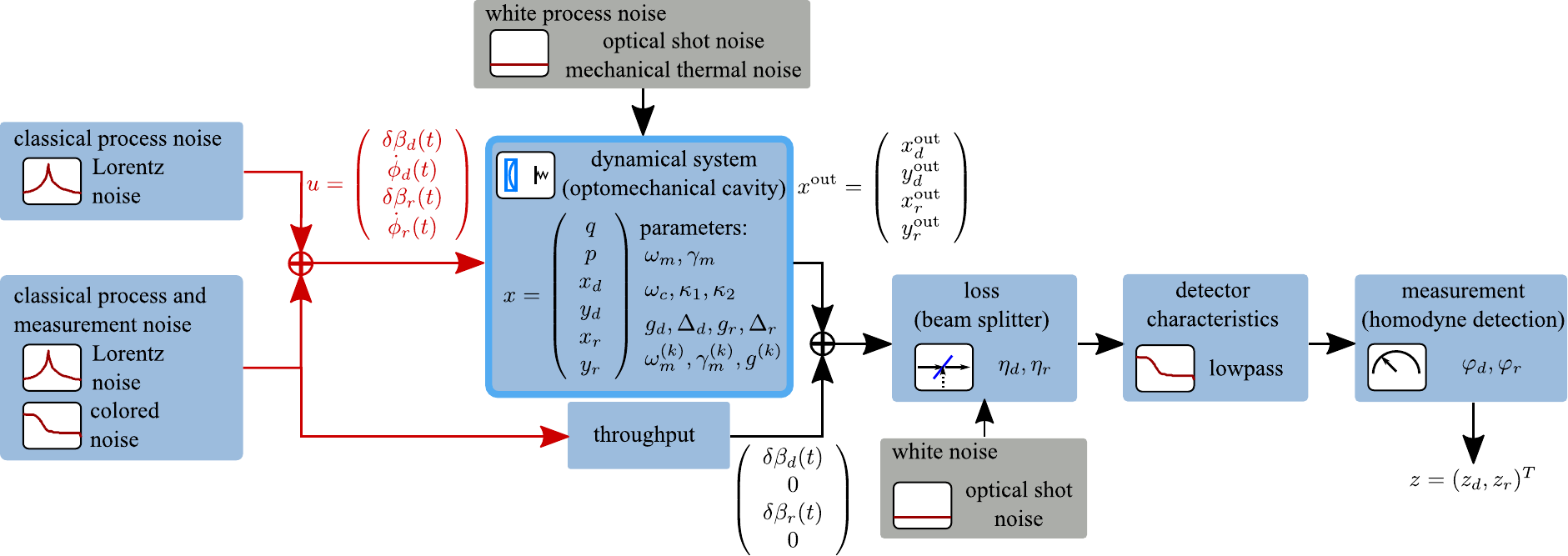}
	\protect\caption{\label{fig:ssmodels} Schematic overview of the complete system model for describing the cavity-optomechanical experiment. State space models of optical noise (Lorentz noise, colored noise), the cavity-optomechanical system, optical loss and homodyne detection are combined via network synthesis to form a single state space model for modeling the total experiment.}
	\par\end{centering}
\end{figure*}

To construct a state space model for the complete optomechanical experiment, we split it into components of independent state space models, which are connected in an appropriate way (using network synthesis, see \cite{gough_series_2009,nurdin_network_2009}). \Fref{fig:ssmodels} shows a schematic of the complete state space model and its  components. These components describe 
\begin{itemize}
	\item the evolution and output of the cavity-optomechanical system, 
	\item optical losses,
	\item homodyne detection,
	\item and various forms of optical noise.
\end{itemize}

\subsection{State evolution}

In our experiment, the cavity-optomechanical system is comprised of two optical cavity modes driven by two external laser fields that interact with a micro-mechanical oscillator, which supports multiple mechanical modes. We are mainly interested in the coupling of the two intra-cavity optical fields to the out-of-plane fundamental mechanical mode of the mechanical oscillator. Then, the state vector consists of the mechanical and optical intra-cavity quadratures $\ts{\vc x}=(q,p,x_{\la},y_{\la},x_{\lb},y_{\lb})^{\trans}$, whose time evolution can be described by the linearized Langevin equations in the rotating, displaced frame:


\begin{subequations}
  \begin{align}
    \dot{q} & =\phantom{-}\om p\\
    \dot{p} & =-\om q-\gamma_mp+\sum_{\mathclap{i={r,d}}}g_i(\cos\theta_i x_i-\sin\theta_iy_i)+\xi\\
    \dot{x}_i & =-\kappa x_i+\Delta_iy_i+g_i\sin\theta_iq+
                \sqrt{2\kappa_1}x_{i,1}^{\mathrm{in}}\nonumber\\
            &\hspace{3.7em}+\sqrt{2\kappa_2}x_{i,2}^{\mathrm{in}}+2\sqrt{\kappa_1}\delta\beta_i+|\alpha_{0,i}|\sin\theta_i\dot{\phi}_i\\
    \dot{y}_i & =-\kappa y_{i}-\Delta_{i}x_{i}+g_{i}\cos\theta_iq+
              \sqrt{2\kappa_1}y_{i,1}^{\mathrm{in}}\nonumber \\
            & \hspace{8,7em}+\sqrt{2\kappa_2}y_{i,2}^{\mathrm{in}}+|\alpha_{0,i}|\cos\theta_i\dot{\phi}_i
  \end{align}
\end{subequations}

The time evolution of the state vector is driven by (white) Brownian thermal noise from the mechanical bath as well as by noise on the driving laser fields, which consists of classical amplitude noise $\db(t)$ and phase noise $\dot{\phi}(t)$, as well as shot noise, which is introduced via the terms $x_{i,j}^{\mathrm{in}},y_{i,j}^{\mathrm{in}}$ with $i\in\{d,r\}$ for the detuned and resonant beam and $j\in\{1,2\}$ for the input coupling mirror and end mirror. 

Shot noise and mechanical thermal noise are white and are incorporated in the state space model directly as process noise $\ts{\vc w}$ and, in the case of measurement shot noise, as measurement noise $\ts{\vc v}$. However, classical laser noise in our experiment is not white. We model it independently as described in \sref{sec:model-color-noise} by constructing state space models, whose output resembles the measured experimental noise characteristics. The output vectors of the noise state space models are treated as deterministic input $\ts{\vc u}=(\db_{\la}(t),\dot{\phi}{}_{\la}(t),\db{}_{\lb}(t),\dot{\phi}{}_{\lb}(t))^{\trans}$, which enter the dynamic and measurement equation of the optomechanical system.

The matrices governing the evolution of the state vector are the process matrix $A\in\mathbb{R}^{6\times6}$, the input matrix $B\in\mathbb{R}^{6\times4}$, the noise matrix $L\in\mathbb{R}^{6\times9}$ and the process noise covariance matrix $W\in\mathbb{R}^{9\times9}$. The process matrix $A$ describes the evolution of the optomechanical system and can be directly read off from the Langevin equations:

\begin{equation}
A=\begin{pmatrix}A_{1}(\om,\gamma_m) & A_{3}(g_{\la},\theta_{\la}) & A_{3}(g_{\lb},\theta_{\lb})\\
A_{4}(g_{\la},\theta_{\la}) & A_{2}(\Delta_{\la}) & 0\\
A_{4}(g_{\lb},\theta_{\lb}) & 0 & A_{2}(\Delta_{\lb})
\end{pmatrix}.\label{eq:Amatrix_OMcavity}
\end{equation}
Note that matrix \eqref{eq:Amatrix_OMcavity} and the following matrices are written in block form. Thus, zeros stand for zero matrices with appropriate dimensions. In \eqref{eq:Amatrix_OMcavity}, the evolution of the mechanical
and optical quadratures is described by the matrices 
\begin{align}
A_{1}(\omega,\gamma) & =\begin{pmatrix}0 & \omega\\
-\omega & -\gamma
\end{pmatrix},\\
A_{2}(\Delta) & =\begin{pmatrix}-\kappa & \Delta\\
-\Delta & -\kappa
\end{pmatrix},
\end{align}
where $\om$ and $\gamma_m$ are the frequency and linewidth (FWHM) of the mechanical mode and $\Delta$ and $\kappa$ the detuning and linewidth (HWHM) of the optical cavity mode, respectively. The interaction between mechanical and optical modes, on the other hand, is described by the matrices 
\begin{align}
A_{3}(g,\theta) & =\begin{pmatrix}0 & 0\\
g\cos{\theta} & -g\sin{\theta}
\end{pmatrix}\\
A_{4}(g,\theta) & =\begin{pmatrix}g\sin{\theta} & 0\\
g\cos{\theta} & 0
\end{pmatrix},
\end{align}
where $g=\sqrt{2}\alpha g_{0}$ is the linearized optomechanical coupling strength and the angle $\theta=\arctan(\Delta/\kappa)$ parametrizes the detuning from cavity resonance. 

To accurately describe the experimental situation we actually have to model multiple mechanical modes of the mechanical oscillator. The generalization of matrix~\eqref{eq:Amatrix_OMcavity} to the multimode case is straightforward: for each additional mechanical mode $k$, one determines the appropriate matrices $A_{1}(\omega_{m}^{(k)},\gamma_{m}^{(k)}),A_{3}(g^{(k)},\theta^{(k)}),A_{4}(g^{(k)},\theta^{(k)})$ from the corresponding optomechanical parameters and adds them as new blocks to \eqref{eq:Amatrix_OMcavity}. Note that adding $M$ additional mechanical modes increases the state space dimension from $6\times6$ to $(6+2M)\times(6+2M)$.

The input matrix $B$ determines the coupling of the classical amplitude and phase noise of the laser to the optical intra-cavity quadratures

\begin{equation}
B=\begin{pmatrix}0 & 0\\
B_1(\theta_{\la},\alpha_{0,\la}) & 0\\
0 & B_1(\theta_{\lb},\alpha_{0,\lb})
\end{pmatrix},\label{eq:Bmatrix_OMcavity}
\end{equation}

with

\begin{equation}
B_1(\theta,\alpha_{0,i})=\begin{pmatrix}\sqrt{2\kappa_{1}} & |\alpha_{0,i}|\sin{\theta}\\
0 & |\alpha_{0,i}|\cos{\theta}
\end{pmatrix},
\end{equation}
where $\kappa_{1}$ is the optical decay rate through the input coupler mirror and $|\alpha_{0,i}|=\sqrt{\frac{2\kappa_1}{\kappa^2+\Delta^2}}\sqrt{\frac{P_i}{\hbar\omega_{0,i}}}$ the intra-cavity photon number. These matrices can again be directly inferred from the Langevin equations, when optical noise is included in their derivation (see, \eg,~\cite{abdi_effect_2011}).

White thermal noise and shotnoise, on the other hand, are treated as process noise $\ts{\vc w}$. Its coupling to the state vector evolution is given by the noise matrix
\begin{equation}
	L=\begin{pmatrix}L_1(\gamma_m) & 0 & 0 & 0 & 0\\
	0 & L_2(\kappa_{1}) & L_2(\kappa_{2}) & 0 & 0\\
	0 & 0 & 0 & L_2(\kappa_{1}) & L_2(\kappa_{2})
\end{pmatrix},\label{eq:Lmatrix_OMcavity}
\end{equation}
with
\begin{equation}
	L_1(\gamma)=-\sqrt{2\gamma}\begin{pmatrix}0\\	1
\end{pmatrix}.
\end{equation}
$L_1(\gamma)$ describes the driving of the mechanical mode by white thermal noise (assumed to act only on the mechanical momentum quadrature) and
\begin{equation}
	L_2(\kappa)=-\sqrt{2\kappa}\openone,
\end{equation}
describes driving of the optical modes by shot noise (with $\openone$ as identity matrix of appropriate dimensions). Note that in our case, shot noise enters both from mirror 1 (input coupler) and mirror 2 (mechanical oscillator) with rates $\kappa_{1}$ and $\kappa_{2}=\kappa-\kappa_{1}$, respectively, where $\kappa_{2}$ also incorporates additional intra-cavity loss. This accounts for the double-sidedness of the cavity.

The process noise covariance matrix is 
\begin{equation}
	W=\mathrm{diag}(n_{m}+\halff,\halff,\dots,\halff),
\end{equation}
where $n_{m}=n_{m}(T,\omega_{m})\approx k_bT/(\hbar\omega_m)$ denotes the thermal occupation of the mechanical bath and $1/2$ is the shot noise contribution of the driving laser fields. 

\subsection{Measurement}

To correctly describe the effect of noise and losses on the measurement, we formally split the measurement into several steps:
\begin{enumerate}
	\item the intra-cavity quadratures $(x_{\la},y_{\la},x_{\lb},y_{\lb})$ are related to the quadratures of the output fields $(\out x_{\la},\out y_{\la},\out x_{\lb},\out y_{\lb})$ via cavity input-output relations; the output quadratures are affected by classical and quantum laser noise,
	\item a state-space model for the optical loss; this takes the output quadratures $(\out x_{\la},\out y_{\la},\out x_{\lb},\out y_{\lb})$ as input and returns the attenuated quadratures $(\damped x_{\la},\damped y_{\la},\damped x_{\lb},\damped y_{\lb})$ that arrive at the homodyne detectors,
	\item a measurement of one (generalized) quadrature per returning laser field, which yields the actual measurements $(z_\la, z_\lb)$.
\end{enumerate}
In conjunction, these steps guarantee that the measurements are physical (\ie~respect uncertainty relations) and that all noise sources are correctly taken into account. Note, we also model the detector spectral response as a digital filter through a separate state space model (see \sref{sec:model-color-noise}).

\subsubsection{Cavity output}

The matrices relating the quadratures of the intra-cavity laser fields to those of the extra-cavity laser fields are the measurement matrix $C\in\mathbb{R}^{4\times6}$, the throughput matrix $D\in\mathbb{R}^{4\times4}$, the measurement noise covariance matrix $V\in\mathbb{R}^{4\times4}$ and the noise cross-correlation matrix $M\in\mathbb{R}^{9\times4}$.

The measurement matrix is
\begin{equation}
	C=\begin{pmatrix}0 & C_1(\kappa_{1}) & 0\\
	0 & 0 & C_1(\kappa_{1})
\end{pmatrix},\label{eq:Cmatrix_OMcavity}
\end{equation}
with 
\begin{equation}
	C_1(\kappa)=\sqrt{2\kappa}\openone.	
\end{equation}
In \eqref{eq:Cmatrix_OMcavity}, $\kappa_{1}$ is the decay rate of the input coupler mirror because we only measure light coming from the input coupler.

Further, the throughput matrix is
\begin{equation}
	D=\begin{pmatrix}D_1 & 0\\
	0 & D_1
\end{pmatrix}\label{eq:Dmatrix_OMcavity}
\end{equation}
with
\begin{equation}
	D_1=\begin{pmatrix}1 & 0\\
	0 & 0
\end{pmatrix}.\label{eq:Dmatrix_OMcavity_singleQuadrature}
\end{equation}
Remember that $D$ in \eqref{eq:SSM_general_measurement} describes the feedthrough of a deterministic input to the state-space model onto the measurement, which is in our case classical laser noise. Hence, $D$ describes the impact of incoming classical laser noise on the extra-cavity quadratures of the returning laser fields.

The fact that \eqref{eq:Dmatrix_OMcavity_singleQuadrature} has an entry only at position $(1,1)$ means that effectively only amplitude noise is reflected off the cavity. In principle, also phase noise $\dot{\phi}_{i}(t)$ is reflected off the cavity; but we can neglect it here since it cancels in homodyne detection (assuming that the local oscillator and reflected optical field have the same instantaneous optical frequency). Note that phase noise does, however, affect the intra-cavity optomechanical state evolution (as described previously).

The measurement noise covariance matrix is
\begin{align}
	V=\mathrm{diag}( & \halff,\halff,\halff,\halff),
\end{align}
\ie, shot noise. Finally, the cross-correlation matrix between process and measurement noise is
\begin{equation}
	M=\begin{pmatrix}0 & 0\\
	M_1 & 0\\
	0 & M_1\\
	0 & 0
\end{pmatrix},
\end{equation}
with 
\begin{equation}
	M_1=\frac{1}{2}\openone.
\end{equation}
The matrix $M$ guarantees that the optical shot noise contribution to the process and measurement noise is perfectly correlated.

\subsubsection{Optical loss}

Optical loss is commonly modeled as a beam splitter with intensity transmission $\eta$ \cite{leonhardt_quantum_1993}, where the signal enters through one port (port $1$) and optical shot noise through the other (port $2$). The state space model of the beam splitter is then only given by the throughput matrix $D\in\mathbb{R}^{4\times4}$, which maps the input quadratures $\vc u=(x_1,y_1,x_2,y_2)^{\trans}$ to the output quadratures $\vc z=(x_3,y_3,x_4,y_4)^{\trans}$:
\begin{equation}
	D=\begin{pmatrix}D_{c}(\tau) & D_{s}(\tau)\\
	-D_{s}(\tau) & D_{c}(\tau)
\end{pmatrix},
\end{equation}
with 
\begin{align}
	D_{c}(\tau) & =\cos\tau\openone,\\
	D_{s}(\tau) & =\sin\tau\openone
\end{align}
and $\tau=\arccos{(\sqrt{\eta})}$. In our case, $(x_1,y_1)$ are the extra-cavity quadratures $(\out x_i,\out y_i)$ and $(x_2,y_2)$ are quadratures of optical shotnoise. The output quadratures are denoted as $(\damped x_{i},\damped y_{i})$.

The optical shotnoise is generated by a separate state space model, which is only made up of the measurement noise covariance matrix
\begin{equation}
	V=\frac{1}{2}\openone.
\end{equation}
Note that each of the two cavity-output beams is fed into a separate loss model with individual loss parameters $\eta_{\la}$ and $\eta_{\lb}$.

\subsubsection{Homodyne detection}

The state space model for homodyne detection guarantees the physicality of the measurement. It maps $(\damped x_{\la},\damped y_{\la},\damped x_{\lb},\damped y_{\lb})$ to the actual physical measurements $\ts{\vc z}$ using the following throughput matrix $\Dt\in\mathbb{R}^{2\times4}$: 
\begin{equation}
	\Dt=\begin{pmatrix}D_1(\varphi_{\la}(t)) & 0\\
	0 & D_1(\varphi_{\lb}(t))
\end{pmatrix},
\end{equation}
with
\begin{equation}
	D_1(\varphi)=\begin{pmatrix}\cos{\varphi} & \sin{\varphi}
\end{pmatrix}.
\end{equation}
The measurement $\ts{\vc z}=(z_{\la}(\varphi_{\la}(t)),z_{\lb}(\varphi_{\lb}(t)))^{\trans}$ can be (explicitly) time-dependent if the phase $\varphi_{i}(t)$ of the local oscillator changes in time.

Note that we do not have to include noise in this final measurement step since all optical noise (shot noise as well as classical noise) has already been included in the quadratures $(\damped x_{\la},\damped y_{\la},\damped x_{\lb},\damped y_{\lb})$.

\section{Kalman filter consistency}
\label{sec:kalm-filt-cons}

\subsection{Statistical consistency tests}

The Kalman filter must be checked for a correct operation in two ways \cite{bar-shalom_estimation_2001}. First, we have to test the filter consistency for a given model, \ie{}, that the estimate $\hat{{\vc x}}$ converges to the true state $\vc x$. In the case of the Kalman filter this means that the sample mean of the estimation error $\ts{\vc\epsilon}=\ts{\vc x}-\ts{\hat{\vc x}}$ goes to zero (which means that the filter is unbiased) and the respective sample covariance matrix converges to $\ts P$. Typically such tests can only be implemented in simulations, as the true state is normally not available in an experiment. Second, we have to ensure that the state-space model accurately describes the experimental system, which means that all experimental parameters have been determined correctly and possible approximations of the system dynamics (including the description of all noise processes) are fulfilled to a sufficient degree. The so-called innovation sequence $\ts{\vc{\nu}}=\ts{\vc z}-\ts C\ts{\hat{{\vc x}}}$---which records the difference between the predicted and the actual measurement results---is our main tool to test the consistency of the Kalman filter with the real experimental situation. For a model perfectly matching the experimental system the innovation sequence is a zero-mean white Gaussian process with a covariance matrix given by $\ts S=\ts C\ts P\ts C^{\trans}+V$ \cite{bar-shalom_estimation_2001,heijden_classification_2005}. For convenience we introduce the normalized process \begin{subequations} 
  \label{eq:SIiv1} 
  \begin{align} 
    \ts{\bar{\vc{\nu}}} & = \ts{L}^{-1}\ts{\vc{\nu}}
  \end{align} 
\end{subequations}
(where $\ts S=\ts L\ts L^{\trans}$ is the Cholesky decomposition of $\ts S$) whose components now have unit variance. Both the whiteness of the process and the correct distribution can be tested for an experimentally recorded innovation sequence. In order to test for whiteness we calculate the so-called periodogram\footnote{The periodogram at Fourier frequency $f_{j}=j/N$ is defined as $I(f_{j})=X_{c}(f_{j})^{2}+X_{s}(f_{j})^{2}$ with the Fourier coefficients $X_{c}(f_{j})=1/\sqrt{N}\sum_{t=1}^{N}\cos{(2\pi f_{j}t)}x_{t}$, $X_{s}(f_{j})=1/\sqrt{N}\sum_{t=1}^{N}\sin{(2\pi f_{j}t)}x_{t}$, $N$ total samples and the random process $x$ with values $x_{t}$ at time $t$. Hence, the periodogram is a sum of independent, squared zero-mean, white Gaussian random vectors and as such asymptotically $\chi^{2}(2)$ distributed with $2$ degrees of freedom.} of a recorded sequence, which asymptotically must be $\chi^{2}$ distributed with two degrees of freedom. The Gaussianity can be tested by calculating the sample's cummulative distribution function (CDF) or the corresponding probability density function (PDF).

\subsection{Consistency test results}

\begin{table}[tb] \begin{center} \begin{tabular}{lcc} \hline\hline & weak coupling & strong coupling \\ \hline sample mean of $\bar{\nu}_\la$ & $0.004$& $-0.012$\\ sample mean of $\bar{\nu}_\lb$ & $0.031$& $0.031$\\ \hline\hline \end{tabular} \end{center} \caption{Mean values of normalized innovations for the weak and strong coupling regime. The expected value is zero.} \label{tab:innov} \end{table}

\begin{figure}[tb] 
\centering{}
\includegraphics[width=1\columnwidth]{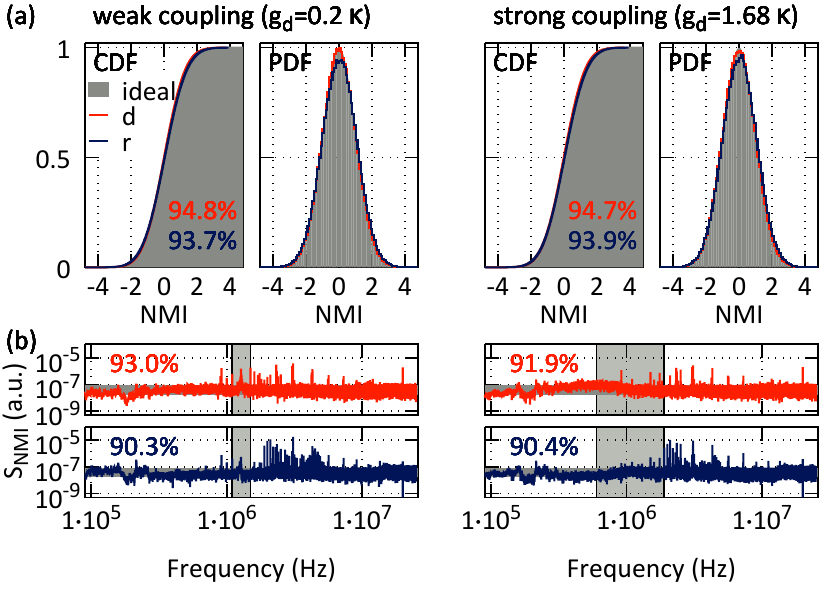} 
\caption{\label{fig:fig_NMI} Statistical analysis of normalized innovation sequence. The cumulative distribution function (CDF) and probability density function (PDF) of the normalized innovations $\bar{\nu}_i$ are shown in (a) for the detuned (red) and resonant (blue) beam, each for the weak (left) and strong (right) coupling regime. The normalized innovations are expected to be a zero-mean Gaussian white noise process with unit variance, \ie{}, $\bar{\nu}_i\sim\mathcal{N}(0,1)$. The numbers in the CDF panel represent the fraction of $\bar{\nu}_i$ that are contained in the expected two-sided $95\%$ confidence region ($\pm2\sigma$) of a random variable distributed as $\mathcal{N}(0,1)$. The noise power spectrum, $S_{\mathrm{NMI}}$, of $\bar{\nu}_i$ is shown in (b). The horizontal, gray shaded area indicates the $95\%$ confidence region of the expected $\chi^2(16)$ distribution. The numbers in the panels reflect the actual value of experimental $\bar{\nu}_i$ contained in this region. The vertical, gray shaded area indicates the region, where the mechanical mode contains most of its noise power.} 
\end{figure}

Here we show the results of the statistical consistency analysis for the data presented in the main text. Table \ref{tab:innov} lists the mean values of the components $\bar{{\nu}}_{\la}$ and $\bar{{\nu}}_{\lb}$ representing the innovation sequence for the measurements of the detuned and resonant beam, respectively. We find that the mean of the normalized innovations is close to zero. To test the distribution of $\bar{\nu}_{\la}$ and $\bar{\nu}_{\lb}$ we calculate their conditional distribution functions and probability density functions, see \fref{fig:fig_NMI}(a). We find that they closely match the expected Gaussian distribution. To be quantitative, we compute the ratio of $\bar{\nu}_{i}$ ($i\in{\la,\lb}$) that are expected to lie within a two-sided $95\%$ confidence region (\ie{}, between $0.025\%$ and $0.975\%$) of the expected Gaussian distribution. For our data, we find values close to the expected $95\%$.

The small deviation can be more closely inspected by considering the noise power spectrum of $\ts{\bar{\nu}}$, which is shown in \fref{fig:fig_NMI}(b). We smooth the periodogram using Welch's method using a Hamming window and splitting the data into eight non-overlapping segments. Then the resulting noise power spectrum is $\chi^{2}$ distributed with $2\cdot8=16$ degrees of freedom. In \fref{fig:fig_NMI}(b) we plot the two-sided $95\%$ confidence region of this distribution as horizontal gray region behind the data. We see that most datapoints lie within this region and thus can be considered white. Most deviations are observed in the low frequency regime around $200\,$kHz (a dip in the noise power) and in the frequency band $2-5\,$MHz (many sharp peaks). Both features are not considered in the model for the Kalman filter and therefore lie outside the $95\%$ confidence region. The series of sharp peaks can be attributed to resonances of a piezoelectric transducer, which is attached to the laser crystal of the driving laser. The spectral dip at $200\,$kHz we attribute to the spectral response of the photo detectors.

Finally we note that we also tested the consistency of the implemented Kalman filter for simulations of a perfectly matched system and find excellent agreement with the criteria presented above.

\section{Real-time feedback}
\label{sec:real-time-feedback}

The execution of the Kalman filter requires calculations of extensive matrix multiplications with a specific dynamic range and minimal delay. Therefore, the implementation of a real-time Kalman filter hinges on a fast and precise hardware platform, which can be based on a high resolution, field programmable gate array (FPGA). Such a hardware platform has to fulfill several criteria. 

First, a minimum precision level has to be achieved. The resolution of the digital calculations needs to be much greater than the minimal step size of the harmonic evolution of an oscillation, \ie, the amplitude of a single phonon. This enables the possible operation at the standard quantum limit. Second, the dynamic range must cover the thermal state of the oscillator. Assuming a typical FPGA clock rate of $500$\,MHz, and the parameters of the mechanical mode of interest, we obtain a necessary resolution of $f_{m}dt/\sqrt{n_{\mathrm{th}}}\approx 4 \cdot 10^{-7} \ll 2^{-27}$. This dynamic range can be covered by a 27\,bit fix-point calculation, as used in the StratixV GS FPGA. This FPGA chip can process 1590 multiplications per clock cycle. It can be used, \eg{}, to calculate a steady-state state space model of one mechanical mode at 1.27\,MHz that is coupled to two optical modes, and incorporates 25 additional mechanical modes and 10 dimensions for modeling colored laser noise. A final criterion concerns the delay due to the detection and the processing of the signals, which needs to be taken into account. We obtain about 50\,ns system delay, when accounting for 4\,m of optical path length and cables, the delay from signal converters and the processing in a 500\,MHz FPGA. This amounts to a phase shift of 23$^{\circ}$ at 1.27\,MHz, the frequency of the mechanical mode of interest. Therefore efficient feedback control of mechanical oscillators with a resonance frequency on the order of 1\,MHz is feasible with state-of-the-art technology.

\clearpage
\pagebreak{}

\section{Symbols and definitions}
\label{sec:symbols-definitions}


\begin{table}[ph]
\begin{tabular}{lllll}
\hline 
symbol  & meaning   \tabularnewline
\hline 
$x_{t}$, $x(t)$  & state vector (random process) \tabularnewline
$z_{t}$  & measurement vector (random process) \tabularnewline
$w_{t}$  & process noise  \tabularnewline
$v_{t}$  & measurement noise  \tabularnewline
$\hat{x}_{t}$, $\hat{x}(t)$  & estimate for $x_{t}$  \tabularnewline
$S_{xx}(\omega)$  & noise power spectrum of $x_{t}$  \tabularnewline
$\mean{\cdot}$  & quantum expectation value  \tabularnewline
$\mathbb{E}[\cdot]$  & classical (population) expectation value  \tabularnewline
 $A_{t}$  & process matrix \tabularnewline
$B_{t}$  & input matrix \tabularnewline
$C_{t}$  & measurement matrix \tabularnewline
$D_{t}$  & throughput matrix \tabularnewline
$W$  & process noise covariance matrix \tabularnewline
$V$  & measurement noise covariance matrix \tabularnewline
$M$  & noise cross-correlation matrix \tabularnewline
$K_{t}$  & Kalman gain matrix \tabularnewline
$P_{t}$  & estimation error covariance matrix \tabularnewline
\hline 
\end{tabular} 
\begin{centering}
\protect\caption{\label{tab:SSsymbols}State space model specific symbols.}
\end{centering}
\end{table}

\begin{table}[ph]
\begin{tabular}{lllc}
\hline 
symbol  & definition  & name  & value \tabularnewline
\hline 
$\omega_{m}$  &  & mechanical frequency  & $2\pi\cdot1.278\cdot10^{6}$\,Hz \tabularnewline
$\gamma_m$  &  & mechanical linewidth (full width at half maximum, FWHM)  & $2\pi\cdot265$\,Hz \tabularnewline
$\kappa$  & $\kappa_{1}+\kappa_{2}$  & cavity decay rate (half width at half maximum, HWHM)  & $0.341\omega_{m}$ \tabularnewline
$\kappa_{1}$  &  & input-coupler cavity decay rate (HWHM)  & $0.2775\omega_{m}$ \tabularnewline
$\kappa_{2}$  &  & total output-coupler and loss cavity decay rate (HWHM)  & $0.0635\omega_{m}$ \tabularnewline
$g_{\la}$  & $\sqrt{2}|\alpha_{0,\la}|g_{0}$  & optomechanical coupling rate detuned beam  & between $0.2\kappa$ and $1.68\kappa$\tabularnewline
$g_{\lb}$  & $\sqrt{2}|\alpha_{0,\lb}|g_{0}$  & optomechanical coupling rate resonant beam  & $0.2\kappa$ \tabularnewline
$g_{0,i}$  & $\frac{\omega_{0,i}}{L}\sqrt{\frac{\hbar}{m_{\mathrm{eff}}\omega_{m}}}$  & single-photon coupling rate  & $2\pi\cdot7.7$\,Hz\tabularnewline
$\theta_{i}$  & $\arctan{\left(\frac{\Delta_{i}}{\kappa}\right)}$  &  & \tabularnewline
$\Delta_{i}$  & $\Delta_{0,i}-\frac{g_{0,i}^{2}|\alpha_{0,i}|^{2}}{\omega_{m}}$  & effective detuning  & $\Delta_{r}=0,\thinspace\Delta_{d}\simeq\om$\tabularnewline
$\Delta_{0,i}$  & $\omega_{c}-\omega_{0,i}$  & cavity-laser detuning  & \tabularnewline
$\omega_{c}$ &  & cavity resonance frequency & \tabularnewline
$\omega_{0,i}$ & $2\pi c/\lambda_i$ & laser frequency & $\sim 1.77\cdot10^{15}$\,Hz\tabularnewline
$P_{i}$ &  & extra-cavity optical power & \tabularnewline
$\beta_{0,i}$  & $\sqrt{\frac{P_{i}}{\hbar\omega_{0}}}$  & extra-cavity amplitude & \tabularnewline
$\delta\beta_{i}(t)$ &   & classical amplitude fluctuation of laser drive (extra-cavity) & \tabularnewline
$\dot{\phi}_{i}(t)$  &  & classical phase noise of laser drive (intra-cavity and extra-cavity) & \tabularnewline
$\alpha_{0,i}$  & $|\alpha_{0,i}|e^{-i\theta_{i}}=\frac{\sqrt{2\kappa_{1}}}{\kappa+i\Delta_{i}}\beta_{0,i}$ & intra-cavity steady-state amplitude  & \tabularnewline
$\delta\alpha_{i}(t)$ &  & classical amplitude fluctuation of laser drive (intra-cavity) & \tabularnewline
$\varphi_{i}$ &  & phase between local oscillator and signal beam in homodyne detection & \tabularnewline
\hline 
\end{tabular}
\begin{centering}
\protect\caption{\label{tab:OMsymbols}\mbox{Optomechanical symbols and definitions. The subscripts $i\in\{\la,\lb\}$ denote the two laser beams.}}
\end{centering}
\end{table}

\twocolumngrid

\newpage
\bibliography{KalmanClean,SMRef}

\begin{thebibliography}{52}%
\makeatletter
\providecommand \@ifxundefined [1]{%
 \@ifx{#1\undefined}
}%
\providecommand \@ifnum [1]{%
 \ifnum #1\expandafter \@firstoftwo
 \else \expandafter \@secondoftwo
 \fi
}%
\providecommand \@ifx [1]{%
 \ifx #1\expandafter \@firstoftwo
 \else \expandafter \@secondoftwo
 \fi
}%
\providecommand \natexlab [1]{#1}%
\providecommand \enquote  [1]{``#1''}%
\providecommand \bibnamefont  [1]{#1}%
\providecommand \bibfnamefont [1]{#1}%
\providecommand \citenamefont [1]{#1}%
\providecommand \href@noop [0]{\@secondoftwo}%
\providecommand \href [0]{\begingroup \@sanitize@url \@href}%
\providecommand \@href[1]{\@@startlink{#1}\@@href}%
\providecommand \@@href[1]{\endgroup#1\@@endlink}%
\providecommand \@sanitize@url [0]{\catcode `\\12\catcode `\$12\catcode
  `\&12\catcode `\#12\catcode `\^12\catcode `\_12\catcode `\%12\relax}%
\providecommand \@@startlink[1]{}%
\providecommand \@@endlink[0]{}%
\providecommand \url  [0]{\begingroup\@sanitize@url \@url }%
\providecommand \@url [1]{\endgroup\@href {#1}{\urlprefix }}%
\providecommand \urlprefix  [0]{URL }%
\providecommand \Eprint [0]{\href }%
\providecommand \doibase [0]{http://dx.doi.org/}%
\providecommand \selectlanguage [0]{\@gobble}%
\providecommand \bibinfo  [0]{\@secondoftwo}%
\providecommand \bibfield  [0]{\@secondoftwo}%
\providecommand \translation [1]{[#1]}%
\providecommand \BibitemOpen [0]{}%
\providecommand \bibitemStop [0]{}%
\providecommand \bibitemNoStop [0]{.\EOS\space}%
\providecommand \EOS [0]{\spacefactor3000\relax}%
\providecommand \BibitemShut  [1]{\csname bibitem#1\endcsname}%
\let\auto@bib@innerbib\@empty
\bibitem [{\citenamefont {Stengel}(1994{\natexlab{a}})}]{Stengel1994}%
  \BibitemOpen
  \bibfield  {author} {\bibinfo {author} {\bibfnamefont {Robert~F}\
  \bibnamefont {Stengel}},\ }\href@noop {} {\emph {\bibinfo {title} {{Optimal
  Control and Estimation}}}}\ (\bibinfo  {publisher} {Dover Publications},\
  \bibinfo {address} {Mineola, NY},\ \bibinfo {year} {1994})\BibitemShut
  {NoStop}%
\bibitem [{\citenamefont {Wiseman}\ and\ \citenamefont
  {Milburn}(2010)}]{Wiseman2010}%
  \BibitemOpen
  \bibfield  {author} {\bibinfo {author} {\bibfnamefont {H}~\bibnamefont
  {Wiseman}}\ and\ \bibinfo {author} {\bibfnamefont {Gerard~J}\ \bibnamefont
  {Milburn}},\ }\href
  {http://www.worldcat.org/title/quantum-measurement-and-control/oclc/434744599\&referer=brief\_results}
  {\emph {\bibinfo {title} {{Quantum Measurement and Control}}}}\ (\bibinfo
  {publisher} {Cambridge University Press},\ \bibinfo {address} {Cambridge},\
  \bibinfo {year} {2010})\BibitemShut {NoStop}%
\bibitem [{\citenamefont {Kalman}(1960)}]{kalman_new_1960}%
  \BibitemOpen
  \bibfield  {author} {\bibinfo {author} {\bibfnamefont {RE}~\bibnamefont
  {Kalman}},\ }\bibfield  {title} {\enquote {\bibinfo {title} {{A New Approach
  to Linear Filtering and Prediction Problems}},}\ }\href
  {http://dx.doi.org/10.1115/1.3662552} {\bibfield  {journal} {\bibinfo
  {journal} {Journal of Fluids Engineering}\ ,\ \bibinfo {pages} {35--45}}
  (\bibinfo {year} {1960})}\BibitemShut {NoStop}%
\bibitem [{\citenamefont {Kalman}\ and\ \citenamefont
  {Bucy}(1961)}]{kalman_new_1961}%
  \BibitemOpen
  \bibfield  {author} {\bibinfo {author} {\bibfnamefont {R~E}\ \bibnamefont
  {Kalman}}\ and\ \bibinfo {author} {\bibfnamefont {R~S}\ \bibnamefont
  {Bucy}},\ }\bibfield  {title} {\enquote {\bibinfo {title} {{New Results in
  Linear Filtering and Prediction Theory}},}\ }\href {\doibase
  10.1115/1.3658902} {\bibfield  {journal} {\bibinfo  {journal} {Journal of
  Fluids Engineering}\ }\textbf {\bibinfo {volume} {83}},\ \bibinfo {pages}
  {95--108} (\bibinfo {year} {1961})}\BibitemShut {NoStop}%
\bibitem [{\citenamefont {Grewal}\ and\ \citenamefont
  {Andrews}(2010)}]{grewal_applications_2010}%
  \BibitemOpen
  \bibfield  {author} {\bibinfo {author} {\bibfnamefont {MS}~\bibnamefont
  {Grewal}}\ and\ \bibinfo {author} {\bibfnamefont {AP}~\bibnamefont
  {Andrews}},\ }\bibfield  {title} {\enquote {\bibinfo {title} {{Applications
  of {Kalman} {Filtering} in {Aerospace} 1960 to the {Present} [{Historical}
  {Perspectives}]}},}\ }\href {\doibase 10.1109/MCS.2010.936465} {\bibfield
  {journal} {\bibinfo  {journal} {IEEE Control Systems}\ }\textbf {\bibinfo
  {volume} {30}},\ \bibinfo {pages} {69--78} (\bibinfo {year}
  {2010})}\BibitemShut {NoStop}%
\bibitem [{\citenamefont {Finn}\ and\ \citenamefont
  {Mukherjee}(2001)}]{Finn2001}%
  \BibitemOpen
  \bibfield  {author} {\bibinfo {author} {\bibfnamefont {Lee~Samuel}\
  \bibnamefont {Finn}}\ and\ \bibinfo {author} {\bibfnamefont {Soma}\
  \bibnamefont {Mukherjee}},\ }\bibfield  {title} {\enquote {\bibinfo {title}
  {{Data conditioning for gravitational wave detectors: A Kalman filter for
  regressing suspension violin modes}},}\ }\href {\doibase
  10.1103/PhysRevD.63.062004} {\bibfield  {journal} {\bibinfo  {journal}
  {Physical Review D}\ }\textbf {\bibinfo {volume} {63}},\ \bibinfo {pages}
  {1--19} (\bibinfo {year} {2001})}\BibitemShut {NoStop}%
\bibitem [{\citenamefont {Geremia}\ \emph {et~al.}(2003)\citenamefont
  {Geremia}, \citenamefont {Stockton}, \citenamefont {Doherty},\ and\
  \citenamefont {Mabuchi}}]{geremia_quantum_2003}%
  \BibitemOpen
  \bibfield  {author} {\bibinfo {author} {\bibfnamefont {JM}~\bibnamefont
  {Geremia}}, \bibinfo {author} {\bibfnamefont {John~K}\ \bibnamefont
  {Stockton}}, \bibinfo {author} {\bibfnamefont {Andrew~C}\ \bibnamefont
  {Doherty}}, \ and\ \bibinfo {author} {\bibfnamefont {Hideo}\ \bibnamefont
  {Mabuchi}},\ }\bibfield  {title} {\enquote {\bibinfo {title} {{Quantum Kalman
  Filtering and the Heisenberg Limit in Atomic Magnetometry}},}\ }\href
  {\doibase 10.1103/PhysRevLett.91.250801} {\bibfield  {journal} {\bibinfo
  {journal} {Physical Review Letters}\ }\textbf {\bibinfo {volume} {91}},\
  \bibinfo {pages} {250801} (\bibinfo {year} {2003})}\BibitemShut {NoStop}%
\bibitem [{\citenamefont {Yonezawa}\ \emph {et~al.}(2012)\citenamefont
  {Yonezawa}, \citenamefont {Nakane}, \citenamefont {Wheatley}, \citenamefont
  {Iwasawa}, \citenamefont {Takeda}, \citenamefont {Arao}, \citenamefont
  {Ohki}, \citenamefont {Tsumura}, \citenamefont {Berry}, \citenamefont
  {Ralph}, \citenamefont {Wiseman}, \citenamefont {Huntington},\ and\
  \citenamefont {Furusawa}}]{yonezawa_quantum-enhanced_2012}%
  \BibitemOpen
  \bibfield  {author} {\bibinfo {author} {\bibfnamefont {Hidehiro}\
  \bibnamefont {Yonezawa}}, \bibinfo {author} {\bibfnamefont {Daisuke}\
  \bibnamefont {Nakane}}, \bibinfo {author} {\bibfnamefont {Trevor~A}\
  \bibnamefont {Wheatley}}, \bibinfo {author} {\bibfnamefont {Kohjiro}\
  \bibnamefont {Iwasawa}}, \bibinfo {author} {\bibfnamefont {Shuntaro}\
  \bibnamefont {Takeda}}, \bibinfo {author} {\bibfnamefont {Hajime}\
  \bibnamefont {Arao}}, \bibinfo {author} {\bibfnamefont {Kentaro}\
  \bibnamefont {Ohki}}, \bibinfo {author} {\bibfnamefont {Koji}\ \bibnamefont
  {Tsumura}}, \bibinfo {author} {\bibfnamefont {Dominic~W}\ \bibnamefont
  {Berry}}, \bibinfo {author} {\bibfnamefont {Timothy~C}\ \bibnamefont
  {Ralph}}, \bibinfo {author} {\bibfnamefont {Howard~M}\ \bibnamefont
  {Wiseman}}, \bibinfo {author} {\bibfnamefont {Elanor~H}\ \bibnamefont
  {Huntington}}, \ and\ \bibinfo {author} {\bibfnamefont {Akira}\ \bibnamefont
  {Furusawa}},\ }\bibfield  {title} {\enquote {\bibinfo {title}
  {{Quantum-Enhanced Optical-Phase Tracking}},}\ }\href {\doibase
  10.1126/science.1225258} {\bibfield  {journal} {\bibinfo  {journal}
  {Science}\ }\textbf {\bibinfo {volume} {337}},\ \bibinfo {pages} {1514--1517}
  (\bibinfo {year} {2012})}\BibitemShut {NoStop}%
\bibitem [{\citenamefont {Tsang}(2009)}]{Tsang2009}%
  \BibitemOpen
  \bibfield  {author} {\bibinfo {author} {\bibfnamefont {Mankei}\ \bibnamefont
  {Tsang}},\ }\bibfield  {title} {\enquote {\bibinfo {title} {{Time-symmetric
  quantum theory of smoothing}},}\ }\href {\doibase
  10.1103/PhysRevLett.102.250403} {\bibfield  {journal} {\bibinfo  {journal}
  {Physical Review Letters}\ }\textbf {\bibinfo {volume} {102}},\ \bibinfo
  {pages} {250403} (\bibinfo {year} {2009})}\BibitemShut {NoStop}%
\bibitem [{\citenamefont {Tsang}\ \emph {et~al.}(2009)\citenamefont {Tsang},
  \citenamefont {Shapiro},\ and\ \citenamefont {Lloyd}}]{Tsang2009a}%
  \BibitemOpen
  \bibfield  {author} {\bibinfo {author} {\bibfnamefont {Mankei}\ \bibnamefont
  {Tsang}}, \bibinfo {author} {\bibfnamefont {Jeffrey~H}\ \bibnamefont
  {Shapiro}}, \ and\ \bibinfo {author} {\bibfnamefont {Seth}\ \bibnamefont
  {Lloyd}},\ }\bibfield  {title} {\enquote {\bibinfo {title} {{Quantum theory
  of optical temporal phase and instantaneous frequency. II. Continuous-time
  limit and state-variable approach to phase-locked loop design}},}\ }\href
  {\doibase 10.1103/PhysRevA.79.053843} {\bibfield  {journal} {\bibinfo
  {journal} {Physical Review A - Atomic, Molecular, and Optical Physics}\
  }\textbf {\bibinfo {volume} {79}},\ \bibinfo {pages} {053843} (\bibinfo
  {year} {2009})}\BibitemShut {NoStop}%
\bibitem [{\citenamefont {Wheatley}\ \emph {et~al.}(2010)\citenamefont
  {Wheatley}, \citenamefont {Berry}, \citenamefont {Yonezawa}, \citenamefont
  {Nakane}, \citenamefont {Arao}, \citenamefont {Pope}, \citenamefont {Ralph},
  \citenamefont {Wiseman}, \citenamefont {Furusawa},\ and\ \citenamefont
  {Huntington}}]{Wheatley2010}%
  \BibitemOpen
  \bibfield  {author} {\bibinfo {author} {\bibfnamefont {T~A}\ \bibnamefont
  {Wheatley}}, \bibinfo {author} {\bibfnamefont {D~W}\ \bibnamefont {Berry}},
  \bibinfo {author} {\bibfnamefont {H}~\bibnamefont {Yonezawa}}, \bibinfo
  {author} {\bibfnamefont {D}~\bibnamefont {Nakane}}, \bibinfo {author}
  {\bibfnamefont {H}~\bibnamefont {Arao}}, \bibinfo {author} {\bibfnamefont
  {D~T}\ \bibnamefont {Pope}}, \bibinfo {author} {\bibfnamefont {T~C}\
  \bibnamefont {Ralph}}, \bibinfo {author} {\bibfnamefont {H~M}\ \bibnamefont
  {Wiseman}}, \bibinfo {author} {\bibfnamefont {A}~\bibnamefont {Furusawa}}, \
  and\ \bibinfo {author} {\bibfnamefont {E~H}\ \bibnamefont {Huntington}},\
  }\bibfield  {title} {\enquote {\bibinfo {title} {{Adaptive optical phase
  estimation using time-symmetric quantum smoothing}},}\ }\href {\doibase
  10.1103/PhysRevLett.104.093601} {\bibfield  {journal} {\bibinfo  {journal}
  {Physical Review Letters}\ }\textbf {\bibinfo {volume} {104}},\ \bibinfo
  {pages} {5--8} (\bibinfo {year} {2010})}\BibitemShut {NoStop}%
\bibitem [{\citenamefont {Braginsky}\ and\ \citenamefont
  {Manukin}(1967)}]{Braginsky1967a}%
  \BibitemOpen
  \bibfield  {author} {\bibinfo {author} {\bibfnamefont {V~B}\ \bibnamefont
  {Braginsky}}\ and\ \bibinfo {author} {\bibfnamefont {A~B}\ \bibnamefont
  {Manukin}},\ }\bibfield  {title} {\enquote {\bibinfo {title} {{Ponderomotive
  effects of electromagnetic radiation}},}\ }\href@noop {} {\bibfield
  {journal} {\bibinfo  {journal} {Soviet Physics JETP}\ }\textbf {\bibinfo
  {volume} {25}},\ \bibinfo {pages} {653--655} (\bibinfo {year}
  {1967})}\BibitemShut {NoStop}%
\bibitem [{\citenamefont {Braginsky}\ and\ \citenamefont
  {Khalili}(1995)}]{braginsky_quantum_1995}%
  \BibitemOpen
  \bibfield  {author} {\bibinfo {author} {\bibfnamefont {Vladimir~B}\
  \bibnamefont {Braginsky}}\ and\ \bibinfo {author} {\bibfnamefont {Farid~Ya}\
  \bibnamefont {Khalili}},\ }\href@noop {} {\emph {\bibinfo {title} {{Quantum
  Measurement}}}}\ (\bibinfo  {publisher} {Cambridge University Press},\
  \bibinfo {year} {1995})\BibitemShut {NoStop}%
\bibitem [{\citenamefont {Aspelmeyer}\ \emph {et~al.}(2014)\citenamefont
  {Aspelmeyer}, \citenamefont {Kippenberg},\ and\ \citenamefont
  {Marquardt}}]{Aspelmeyer2014}%
  \BibitemOpen
  \bibfield  {author} {\bibinfo {author} {\bibfnamefont {Markus}\ \bibnamefont
  {Aspelmeyer}}, \bibinfo {author} {\bibfnamefont {Tobias~J}\ \bibnamefont
  {Kippenberg}}, \ and\ \bibinfo {author} {\bibfnamefont {Florian}\
  \bibnamefont {Marquardt}},\ }\bibfield  {title} {\enquote {\bibinfo {title}
  {{Cavity optomechanics}},}\ }\href {\doibase
  http://dx.doi.org/10.1103/RevModPhys.86.1391} {\bibfield  {journal} {\bibinfo
   {journal} {Reviews of Modern Physics}\ }\textbf {\bibinfo {volume} {86}},\
  \bibinfo {pages} {1391} (\bibinfo {year} {2014})}\BibitemShut {NoStop}%
\bibitem [{\citenamefont {Paternostro}\ \emph {et~al.}(2006)\citenamefont
  {Paternostro}, \citenamefont {Gigan}, \citenamefont {Kim}, \citenamefont
  {Blaser}, \citenamefont {B\"{o}hm},\ and\ \citenamefont
  {Aspelmeyer}}]{Paternostro2006}%
  \BibitemOpen
  \bibfield  {author} {\bibinfo {author} {\bibfnamefont {M}~\bibnamefont
  {Paternostro}}, \bibinfo {author} {\bibfnamefont {S}~\bibnamefont {Gigan}},
  \bibinfo {author} {\bibfnamefont {M~S}\ \bibnamefont {Kim}}, \bibinfo
  {author} {\bibfnamefont {F}~\bibnamefont {Blaser}}, \bibinfo {author}
  {\bibfnamefont {H~R}\ \bibnamefont {B\"{o}hm}}, \ and\ \bibinfo {author}
  {\bibfnamefont {M}~\bibnamefont {Aspelmeyer}},\ }\bibfield  {title} {\enquote
  {\bibinfo {title} {{Reconstructing the dynamics of a movable mirror in a
  detuned optical cavity}},}\ }\href@noop {} {\bibfield  {journal} {\bibinfo
  {journal} {New Journal of Physics}\ }\textbf {\bibinfo {volume} {8}},\
  \bibinfo {pages} {107} (\bibinfo {year} {2006})}\BibitemShut {NoStop}%
\bibitem [{\citenamefont {Palomaki}\ \emph {et~al.}(2013)\citenamefont
  {Palomaki}, \citenamefont {Harlow}, \citenamefont {Teufel}, \citenamefont
  {Simmonds},\ and\ \citenamefont {Lehnert}}]{Palomaki2013}%
  \BibitemOpen
  \bibfield  {author} {\bibinfo {author} {\bibfnamefont {T~A}\ \bibnamefont
  {Palomaki}}, \bibinfo {author} {\bibfnamefont {J~W}\ \bibnamefont {Harlow}},
  \bibinfo {author} {\bibfnamefont {J~D}\ \bibnamefont {Teufel}}, \bibinfo
  {author} {\bibfnamefont {R~W}\ \bibnamefont {Simmonds}}, \ and\ \bibinfo
  {author} {\bibfnamefont {K~W}\ \bibnamefont {Lehnert}},\ }\bibfield  {title}
  {\enquote {\bibinfo {title} {{Coherent state transfer between itinerant
  microwave fields and a mechanical oscillator.}}}\ }\href {\doibase
  10.1038/nature11915} {\bibfield  {journal} {\bibinfo  {journal} {Nature}\
  }\textbf {\bibinfo {volume} {495}},\ \bibinfo {pages} {210--4} (\bibinfo
  {year} {2013})}\BibitemShut {NoStop}%
\bibitem [{\citenamefont {Rugar}\ and\ \citenamefont
  {Gr\"{u}tter}(1991)}]{Rugar1991}%
  \BibitemOpen
  \bibfield  {author} {\bibinfo {author} {\bibfnamefont {D}~\bibnamefont
  {Rugar}}\ and\ \bibinfo {author} {\bibfnamefont {P}~\bibnamefont
  {Gr\"{u}tter}},\ }\bibfield  {title} {\enquote {\bibinfo {title} {{Mechanical
  parametric amplification and thermomechanical noise squeezing}},}\ }\href
  {\doibase 10.1103/PhysRevLett.67.699} {\bibfield  {journal} {\bibinfo
  {journal} {Physical Review Letters}\ }\textbf {\bibinfo {volume} {67}},\
  \bibinfo {pages} {699--702} (\bibinfo {year} {1991})}\BibitemShut {NoStop}%
\bibitem [{\citenamefont {Hadjar}\ \emph {et~al.}(1999)\citenamefont {Hadjar},
  \citenamefont {Cohadon}, \citenamefont {Aminoff}, \citenamefont {Pinard},\
  and\ \citenamefont {Heidmann}}]{Hadjar1999}%
  \BibitemOpen
  \bibfield  {author} {\bibinfo {author} {\bibfnamefont {Y}~\bibnamefont
  {Hadjar}}, \bibinfo {author} {\bibfnamefont {P~F}\ \bibnamefont {Cohadon}},
  \bibinfo {author} {\bibfnamefont {C~G}\ \bibnamefont {Aminoff}}, \bibinfo
  {author} {\bibfnamefont {M}~\bibnamefont {Pinard}}, \ and\ \bibinfo {author}
  {\bibfnamefont {A}~\bibnamefont {Heidmann}},\ }\bibfield  {title} {\enquote
  {\bibinfo {title} {{High-sensitivity optical measurement of mechanical
  Brownian motion}},}\ }\href {\doibase 10.1209/epl/i1999-00422-6} {\bibfield
  {journal} {\bibinfo  {journal} {Europhysics Letters (EPL)}\ }\textbf
  {\bibinfo {volume} {47}},\ \bibinfo {pages} {545--551} (\bibinfo {year}
  {1999})}\BibitemShut {NoStop}%
\bibitem [{\citenamefont {Briant}\ \emph {et~al.}(2003)\citenamefont {Briant},
  \citenamefont {Cohadon}, \citenamefont {Pinard},\ and\ \citenamefont
  {Heidmann}}]{Briant2003a}%
  \BibitemOpen
  \bibfield  {author} {\bibinfo {author} {\bibfnamefont {T}~\bibnamefont
  {Briant}}, \bibinfo {author} {\bibfnamefont {P-F}\ \bibnamefont {Cohadon}},
  \bibinfo {author} {\bibfnamefont {M}~\bibnamefont {Pinard}}, \ and\ \bibinfo
  {author} {\bibfnamefont {A}~\bibnamefont {Heidmann}},\ }\bibfield  {title}
  {\enquote {\bibinfo {title} {{Optical phase-space reconstruction of mirror
  motion at the attometer level}},}\ }\href@noop {} {\bibfield  {journal}
  {\bibinfo  {journal} {The European Physical Journal D}\ }\textbf {\bibinfo
  {volume} {22}},\ \bibinfo {pages} {131--140} (\bibinfo {year}
  {2003})}\BibitemShut {NoStop}%
\bibitem [{\citenamefont {Iwasawa}\ \emph {et~al.}(2013)\citenamefont
  {Iwasawa}, \citenamefont {Makino}, \citenamefont {Yonezawa}, \citenamefont
  {Tsang}, \citenamefont {Davidovic}, \citenamefont {Huntington},\ and\
  \citenamefont {Furusawa}}]{Iwasawa2013}%
  \BibitemOpen
  \bibfield  {author} {\bibinfo {author} {\bibfnamefont {Kohjiro}\ \bibnamefont
  {Iwasawa}}, \bibinfo {author} {\bibfnamefont {Kenzo}\ \bibnamefont {Makino}},
  \bibinfo {author} {\bibfnamefont {Hidehiro}\ \bibnamefont {Yonezawa}},
  \bibinfo {author} {\bibfnamefont {Mankei}\ \bibnamefont {Tsang}}, \bibinfo
  {author} {\bibfnamefont {Aleksandar}\ \bibnamefont {Davidovic}}, \bibinfo
  {author} {\bibfnamefont {Elanor}\ \bibnamefont {Huntington}}, \ and\ \bibinfo
  {author} {\bibfnamefont {Akira}\ \bibnamefont {Furusawa}},\ }\bibfield
  {title} {\enquote {\bibinfo {title} {{Quantum-Limited Mirror-Motion
  Estimation}},}\ }\href {\doibase 10.1103/PhysRevLett.111.163602} {\bibfield
  {journal} {\bibinfo  {journal} {Physical Review Letters}\ }\textbf {\bibinfo
  {volume} {111}},\ \bibinfo {pages} {163602} (\bibinfo {year}
  {2013})}\BibitemShut {NoStop}%
\bibitem [{\citenamefont {Wilson}\ \emph {et~al.}(2014)\citenamefont {Wilson},
  \citenamefont {Sudhir}, \citenamefont {Piro}, \citenamefont {Schilling},
  \citenamefont {Ghadimi},\ and\ \citenamefont {Kippenberg}}]{Wilson2014}%
  \BibitemOpen
  \bibfield  {author} {\bibinfo {author} {\bibfnamefont {D~J}\ \bibnamefont
  {Wilson}}, \bibinfo {author} {\bibfnamefont {V}~\bibnamefont {Sudhir}},
  \bibinfo {author} {\bibfnamefont {N}~\bibnamefont {Piro}}, \bibinfo {author}
  {\bibfnamefont {R}~\bibnamefont {Schilling}}, \bibinfo {author}
  {\bibfnamefont {A}~\bibnamefont {Ghadimi}}, \ and\ \bibinfo {author}
  {\bibfnamefont {T~J}\ \bibnamefont {Kippenberg}},\ }\bibfield  {title}
  {\enquote {\bibinfo {title} {{Measurement and control of a mechanical
  oscillator at its thermal decoherence rate}},}\ }\href@noop {} {\  (\bibinfo
  {year} {2014})},\ \Eprint {http://arxiv.org/abs/arXiv:1410.6191v2}
  {arXiv:1410.6191v2} \BibitemShut {NoStop}%
\bibitem [{\citenamefont {Carmichael}(1993)}]{carmichael_open_1993}%
  \BibitemOpen
  \bibfield  {author} {\bibinfo {author} {\bibfnamefont {Howard}\ \bibnamefont
  {Carmichael}},\ }\href@noop {} {\emph {\bibinfo {title} {{An open systems
  approach to quantum optics}}}}\ (\bibinfo  {publisher} {Springer-Verlag},\
  \bibinfo {year} {1993})\BibitemShut {NoStop}%
\bibitem [{\citenamefont {Gardiner}\ and\ \citenamefont
  {Zoller}(2004)}]{gardiner_quantum_2004}%
  \BibitemOpen
  \bibfield  {author} {\bibinfo {author} {\bibfnamefont {Crispin~W}\
  \bibnamefont {Gardiner}}\ and\ \bibinfo {author} {\bibfnamefont {Peter}\
  \bibnamefont {Zoller}},\ }\href@noop {} {\emph {\bibinfo {title} {{Quantum
  noise}}}},\ \bibinfo {edition} {3rd}\ ed.\ (\bibinfo  {publisher}
  {Springer},\ \bibinfo {year} {2004})\BibitemShut {NoStop}%
\bibitem [{\citenamefont {Bouten}\ \emph {et~al.}(2007)\citenamefont {Bouten},
  \citenamefont {Van~Handel},\ and\ \citenamefont
  {James}}]{bouten_introduction_2007}%
  \BibitemOpen
  \bibfield  {author} {\bibinfo {author} {\bibfnamefont {Luc}\ \bibnamefont
  {Bouten}}, \bibinfo {author} {\bibfnamefont {Ramon}\ \bibnamefont
  {Van~Handel}}, \ and\ \bibinfo {author} {\bibfnamefont {Matthew~R}\
  \bibnamefont {James}},\ }\bibfield  {title} {\enquote {\bibinfo {title} {{An
  Introduction to Quantum Filtering}},}\ }\href {\doibase 10.1137/060651239}
  {\bibfield  {journal} {\bibinfo  {journal} {{SIAM} Journal on Control and
  Optimization}\ }\textbf {\bibinfo {volume} {46}},\ \bibinfo {pages}
  {2199--2241} (\bibinfo {year} {2007})}\BibitemShut {NoStop}%
\bibitem [{\citenamefont {Mancini}\ \emph {et~al.}(1998)\citenamefont
  {Mancini}, \citenamefont {Vitali},\ and\ \citenamefont
  {Tombesi}}]{mancini_optomechanical_1998}%
  \BibitemOpen
  \bibfield  {author} {\bibinfo {author} {\bibfnamefont {Stefano}\ \bibnamefont
  {Mancini}}, \bibinfo {author} {\bibfnamefont {David}\ \bibnamefont {Vitali}},
  \ and\ \bibinfo {author} {\bibfnamefont {Paolo}\ \bibnamefont {Tombesi}},\
  }\bibfield  {title} {\enquote {\bibinfo {title} {{Optomechanical Cooling of a
  Macroscopic Oscillator by Homodyne Feedback}},}\ }\href {\doibase
  10.1103/PhysRevLett.80.688} {\bibfield  {journal} {\bibinfo  {journal}
  {Physical Review Letters}\ }\textbf {\bibinfo {volume} {80}},\ \bibinfo
  {pages} {688--691} (\bibinfo {year} {1998})}\BibitemShut {NoStop}%
\bibitem [{\citenamefont {Doherty}\ and\ \citenamefont
  {Jacobs}(1999)}]{doherty_feedback_1999}%
  \BibitemOpen
  \bibfield  {author} {\bibinfo {author} {\bibfnamefont {A~C}\ \bibnamefont
  {Doherty}}\ and\ \bibinfo {author} {\bibfnamefont {K}~\bibnamefont
  {Jacobs}},\ }\bibfield  {title} {\enquote {\bibinfo {title} {{Feedback
  control of quantum systems using continuous state estimation}},}\ }\href
  {\doibase 10.1103/PhysRevA.60.2700} {\bibfield  {journal} {\bibinfo
  {journal} {Physical Review A}\ }\textbf {\bibinfo {volume} {60}},\ \bibinfo
  {pages} {2700--2711} (\bibinfo {year} {1999})}\BibitemShut {NoStop}%
\bibitem [{\citenamefont {Hopkins}\ \emph {et~al.}(2003)\citenamefont
  {Hopkins}, \citenamefont {Jacobs}, \citenamefont {Habib},\ and\ \citenamefont
  {Schwab}}]{hopkins_feedback_2003}%
  \BibitemOpen
  \bibfield  {author} {\bibinfo {author} {\bibfnamefont {Asa}\ \bibnamefont
  {Hopkins}}, \bibinfo {author} {\bibfnamefont {Kurt}\ \bibnamefont {Jacobs}},
  \bibinfo {author} {\bibfnamefont {Salman}\ \bibnamefont {Habib}}, \ and\
  \bibinfo {author} {\bibfnamefont {Keith}\ \bibnamefont {Schwab}},\ }\bibfield
   {title} {\enquote {\bibinfo {title} {{Feedback cooling of a nanomechanical
  resonator}},}\ }\href {\doibase 10.1103/PhysRevB.68.235328} {\bibfield
  {journal} {\bibinfo  {journal} {Physical Review B}\ }\textbf {\bibinfo
  {volume} {68}},\ \bibinfo {pages} {235328} (\bibinfo {year}
  {2003})}\BibitemShut {NoStop}%
\bibitem [{\citenamefont {Danilishin}\ \emph {et~al.}(2008)\citenamefont
  {Danilishin}, \citenamefont {Müller-Ebhardt}, \citenamefont {Rehbein},
  \citenamefont {Somiya}, \citenamefont {Schnabel}, \citenamefont {Danzmann},
  \citenamefont {Corbitt}, \citenamefont {Wipf}, \citenamefont {Mavalvala},\
  and\ \citenamefont {Chen}}]{danilishin_creation_2008}%
  \BibitemOpen
  \bibfield  {author} {\bibinfo {author} {\bibfnamefont {Stefan}\ \bibnamefont
  {Danilishin}}, \bibinfo {author} {\bibfnamefont {Helge}\ \bibnamefont
  {Müller-Ebhardt}}, \bibinfo {author} {\bibfnamefont {Henning}\ \bibnamefont
  {Rehbein}}, \bibinfo {author} {\bibfnamefont {Kentaro}\ \bibnamefont
  {Somiya}}, \bibinfo {author} {\bibfnamefont {Roman}\ \bibnamefont
  {Schnabel}}, \bibinfo {author} {\bibfnamefont {Karsten}\ \bibnamefont
  {Danzmann}}, \bibinfo {author} {\bibfnamefont {Thomas}\ \bibnamefont
  {Corbitt}}, \bibinfo {author} {\bibfnamefont {Christopher}\ \bibnamefont
  {Wipf}}, \bibinfo {author} {\bibfnamefont {Nergis}\ \bibnamefont
  {Mavalvala}}, \ and\ \bibinfo {author} {\bibfnamefont {Yanbei}\ \bibnamefont
  {Chen}},\ }\bibfield  {title} {\enquote {\bibinfo {title} {{Creation of a
  quantum oscillator by classical control}},}\ }\href
  {http://arxiv.org/abs/0809.2024} {\bibfield  {journal} {\bibinfo  {journal}
  {0809.2024}\ } (\bibinfo {year} {2008})}\BibitemShut {NoStop}%
\bibitem [{\citenamefont {Müller-Ebhardt}\ \emph {et~al.}(2009)\citenamefont
  {Müller-Ebhardt}, \citenamefont {Rehbein}, \citenamefont {Li}, \citenamefont
  {Mino}, \citenamefont {Somiya}, \citenamefont {Schnabel}, \citenamefont
  {Danzmann},\ and\ \citenamefont {Chen}}]{muller-ebhardt_quantum-state_2009}%
  \BibitemOpen
  \bibfield  {author} {\bibinfo {author} {\bibfnamefont {Helge}\ \bibnamefont
  {Müller-Ebhardt}}, \bibinfo {author} {\bibfnamefont {Henning}\ \bibnamefont
  {Rehbein}}, \bibinfo {author} {\bibfnamefont {Chao}\ \bibnamefont {Li}},
  \bibinfo {author} {\bibfnamefont {Yasushi}\ \bibnamefont {Mino}}, \bibinfo
  {author} {\bibfnamefont {Kentaro}\ \bibnamefont {Somiya}}, \bibinfo {author}
  {\bibfnamefont {Roman}\ \bibnamefont {Schnabel}}, \bibinfo {author}
  {\bibfnamefont {Karsten}\ \bibnamefont {Danzmann}}, \ and\ \bibinfo {author}
  {\bibfnamefont {Yanbei}\ \bibnamefont {Chen}},\ }\bibfield  {title} {\enquote
  {\bibinfo {title} {{Quantum-state preparation and macroscopic entanglement in
  gravitational-wave detectors}},}\ }\href {\doibase
  10.1103/PhysRevA.80.043802} {\bibfield  {journal} {\bibinfo  {journal}
  {Physical Review A}\ }\textbf {\bibinfo {volume} {80}},\ \bibinfo {pages}
  {043802} (\bibinfo {year} {2009})}\BibitemShut {NoStop}%
\bibitem [{\citenamefont {Belavkin}(1980)}]{belavkin_optimal_1980}%
  \BibitemOpen
  \bibfield  {author} {\bibinfo {author} {\bibfnamefont {VP}~\bibnamefont
  {Belavkin}},\ }\bibfield  {title} {\enquote {\bibinfo {title} {{Optimal
  filtering of Markov signals with quantum white noise}},}\ }\href@noop {}
  {\bibfield  {journal} {\bibinfo  {journal} {Radio Eng Electron Physics}\
  }\textbf {\bibinfo {volume} {25}},\ \bibinfo {pages} {1445--1453} (\bibinfo
  {year} {1980})}\BibitemShut {NoStop}%
\bibitem [{\citenamefont {Hofer}\ and\ \citenamefont
  {Hammerer}(2015)}]{hofer_entanglement-enhanced_2015}%
  \BibitemOpen
  \bibfield  {author} {\bibinfo {author} {\bibfnamefont {Sebastian~G.}\
  \bibnamefont {Hofer}}\ and\ \bibinfo {author} {\bibfnamefont {Klemens}\
  \bibnamefont {Hammerer}},\ }\bibfield  {title} {\enquote {\bibinfo {title}
  {Entanglement-enhanced time-continuous quantum control in optomechanics},}\
  }\href {\doibase 10.1103/PhysRevA.91.033822} {\bibfield  {journal} {\bibinfo
  {journal} {Physical Review A}\ }\textbf {\bibinfo {volume} {91}},\ \bibinfo
  {pages} {033822} (\bibinfo {year} {2015})}\BibitemShut {NoStop}%
\bibitem [{\citenamefont {Giovannetti}\ and\ \citenamefont
  {Vitali}(2001)}]{giovannetti_phase-noise_2001}%
  \BibitemOpen
  \bibfield  {author} {\bibinfo {author} {\bibfnamefont {Vittorio}\
  \bibnamefont {Giovannetti}}\ and\ \bibinfo {author} {\bibfnamefont {David}\
  \bibnamefont {Vitali}},\ }\bibfield  {title} {\enquote {\bibinfo {title}
  {{Phase-noise measurement in a cavity with a movable mirror undergoing
  quantum Brownian motion}},}\ }\href {\doibase 10.1103/PhysRevA.63.023812}
  {\bibfield  {journal} {\bibinfo  {journal} {Physical Review A}\ }\textbf
  {\bibinfo {volume} {63}},\ \bibinfo {pages} {023812} (\bibinfo {year}
  {2001})}\BibitemShut {NoStop}%
\bibitem [{\citenamefont {Rabl}\ \emph {et~al.}(2009)\citenamefont {Rabl},
  \citenamefont {Genes}, \citenamefont {Hammerer},\ and\ \citenamefont
  {Aspelmeyer}}]{rabl_phase-noise_2009}%
  \BibitemOpen
  \bibfield  {author} {\bibinfo {author} {\bibfnamefont {P}~\bibnamefont
  {Rabl}}, \bibinfo {author} {\bibfnamefont {C}~\bibnamefont {Genes}}, \bibinfo
  {author} {\bibfnamefont {K}~\bibnamefont {Hammerer}}, \ and\ \bibinfo
  {author} {\bibfnamefont {M}~\bibnamefont {Aspelmeyer}},\ }\bibfield  {title}
  {\enquote {\bibinfo {title} {{Phase-noise induced limitations on cooling and
  coherent evolution in optomechanical systems}},}\ }\href {\doibase
  10.1103/PhysRevA.80.063819} {\bibfield  {journal} {\bibinfo  {journal}
  {Physical Review A}\ }\textbf {\bibinfo {volume} {80}},\ \bibinfo {pages}
  {063819} (\bibinfo {year} {2009})}\BibitemShut {NoStop}%
\bibitem [{\citenamefont {Abdi}\ \emph {et~al.}(2011)\citenamefont {Abdi},
  \citenamefont {Barzanjeh}, \citenamefont {Tombesi},\ and\ \citenamefont
  {Vitali}}]{abdi_effect_2011}%
  \BibitemOpen
  \bibfield  {author} {\bibinfo {author} {\bibfnamefont {M}~\bibnamefont
  {Abdi}}, \bibinfo {author} {\bibfnamefont {Sh}~\bibnamefont {Barzanjeh}},
  \bibinfo {author} {\bibfnamefont {P}~\bibnamefont {Tombesi}}, \ and\ \bibinfo
  {author} {\bibfnamefont {D}~\bibnamefont {Vitali}},\ }\bibfield  {title}
  {\enquote {\bibinfo {title} {{Effect of phase noise on the generation of
  stationary entanglement in cavity optomechanics}},}\ }\href {\doibase
  10.1103/PhysRevA.84.032325} {\bibfield  {journal} {\bibinfo  {journal}
  {Physical Review A}\ }\textbf {\bibinfo {volume} {84}},\ \bibinfo {pages}
  {032325} (\bibinfo {year} {2011})}\BibitemShut {NoStop}%
\bibitem [{\citenamefont {Ghobadi}\ \emph {et~al.}(2011)\citenamefont
  {Ghobadi}, \citenamefont {Bahrampour},\ and\ \citenamefont
  {Simon}}]{ghobadi_optomechanical_2011}%
  \BibitemOpen
  \bibfield  {author} {\bibinfo {author} {\bibfnamefont {R}~\bibnamefont
  {Ghobadi}}, \bibinfo {author} {\bibfnamefont {A~R}\ \bibnamefont
  {Bahrampour}}, \ and\ \bibinfo {author} {\bibfnamefont {C}~\bibnamefont
  {Simon}},\ }\bibfield  {title} {\enquote {\bibinfo {title} {{Optomechanical
  entanglement in the presence of laser phase noise}},}\ }\href {\doibase
  10.1103/PhysRevA.84.063827} {\bibfield  {journal} {\bibinfo  {journal}
  {Physical Review A}\ }\textbf {\bibinfo {volume} {84}},\ \bibinfo {pages}
  {063827} (\bibinfo {year} {2011})}\BibitemShut {NoStop}%
\bibitem [{\citenamefont {Genes}\ \emph {et~al.}(2008)\citenamefont {Genes},
  \citenamefont {Vitali}, \citenamefont {Tombesi}, \citenamefont {Gigan},\ and\
  \citenamefont {Aspelmeyer}}]{genes_ground-state_2008}%
  \BibitemOpen
  \bibfield  {author} {\bibinfo {author} {\bibfnamefont {C}~\bibnamefont
  {Genes}}, \bibinfo {author} {\bibfnamefont {D}~\bibnamefont {Vitali}},
  \bibinfo {author} {\bibfnamefont {P}~\bibnamefont {Tombesi}}, \bibinfo
  {author} {\bibfnamefont {S}~\bibnamefont {Gigan}}, \ and\ \bibinfo {author}
  {\bibfnamefont {M}~\bibnamefont {Aspelmeyer}},\ }\bibfield  {title} {\enquote
  {\bibinfo {title} {{Ground-state cooling of a micromechanical oscillator:
  Comparing cold damping and cavity-assisted cooling schemes}},}\ }\href
  {\doibase 10.1103/PhysRevA.77.033804} {\bibfield  {journal} {\bibinfo
  {journal} {Physical Review A}\ }\textbf {\bibinfo {volume} {77}},\ \bibinfo
  {pages} {033804--9} (\bibinfo {year} {2008})}\BibitemShut {NoStop}%
\bibitem [{\citenamefont {Brennecke}\ \emph {et~al.}(2008)\citenamefont
  {Brennecke}, \citenamefont {Ritter}, \citenamefont {Donner},\ and\
  \citenamefont {Esslinger}}]{brennecke_2008}%
  \BibitemOpen
  \bibfield  {author} {\bibinfo {author} {\bibfnamefont {Ferdinand}\
  \bibnamefont {Brennecke}}, \bibinfo {author} {\bibfnamefont {Stephan}\
  \bibnamefont {Ritter}}, \bibinfo {author} {\bibfnamefont {Tobias}\
  \bibnamefont {Donner}}, \ and\ \bibinfo {author} {\bibfnamefont {Tilman}\
  \bibnamefont {Esslinger}},\ }\bibfield  {title} {\enquote {\bibinfo {title}
  {Cavity optomechanics with a bose-einstein condensate},}\ }\href {\doibase
  10.1126/science.1163218} {\bibfield  {journal} {\bibinfo  {journal}
  {Science}\ }\textbf {\bibinfo {volume} {322}},\ \bibinfo {pages} {235--238}
  (\bibinfo {year} {2008})}\BibitemShut {NoStop}%
\bibitem [{\citenamefont {Murch}\ \emph {et~al.}(2008)\citenamefont {Murch},
  \citenamefont {Moore}, \citenamefont {Gupta},\ and\ \citenamefont
  {Stamper-Kurn}}]{Murch_observation_2008}%
  \BibitemOpen
  \bibfield  {author} {\bibinfo {author} {\bibfnamefont {Kater~W}\ \bibnamefont
  {Murch}}, \bibinfo {author} {\bibfnamefont {Kevin~L}\ \bibnamefont {Moore}},
  \bibinfo {author} {\bibfnamefont {Subhadeep}\ \bibnamefont {Gupta}}, \ and\
  \bibinfo {author} {\bibfnamefont {Dan~M}\ \bibnamefont {Stamper-Kurn}},\
  }\bibfield  {title} {\enquote {\bibinfo {title} {{Observation of
  quantum-measurement backaction with an ultracold atomic gas}},}\ }\href
  {\doibase 10.1038/nphys965} {\bibfield  {journal} {\bibinfo  {journal}
  {Nature Physics}\ }\textbf {\bibinfo {volume} {4}},\ \bibinfo {pages}
  {561--564} (\bibinfo {year} {2008})}\BibitemShut {NoStop}%
\bibitem [{\citenamefont {Purdy}\ \emph
  {et~al.}(2013{\natexlab{a}})\citenamefont {Purdy}, \citenamefont {Peterson},\
  and\ \citenamefont {Regal}}]{purdy_observation_2013}%
  \BibitemOpen
  \bibfield  {author} {\bibinfo {author} {\bibfnamefont {T~P}\ \bibnamefont
  {Purdy}}, \bibinfo {author} {\bibfnamefont {R~W}\ \bibnamefont {Peterson}}, \
  and\ \bibinfo {author} {\bibfnamefont {C~A}\ \bibnamefont {Regal}},\
  }\bibfield  {title} {\enquote {\bibinfo {title} {{Observation of Radiation
  Pressure Shot Noise on a Macroscopic Object}},}\ }\href {\doibase
  10.1126/science.1231282} {\bibfield  {journal} {\bibinfo  {journal}
  {Science}\ }\textbf {\bibinfo {volume} {339}},\ \bibinfo {pages} {801--804}
  (\bibinfo {year} {2013}{\natexlab{a}})}\BibitemShut {NoStop}%
\bibitem [{\citenamefont {Müller-Ebhardt}\ \emph {et~al.}(2008)\citenamefont
  {Müller-Ebhardt}, \citenamefont {Rehbein}, \citenamefont {Schnabel},
  \citenamefont {Danzmann},\ and\ \citenamefont
  {Chen}}]{muller-ebhardt_entanglement_2008}%
  \BibitemOpen
  \bibfield  {author} {\bibinfo {author} {\bibfnamefont {Helge}\ \bibnamefont
  {Müller-Ebhardt}}, \bibinfo {author} {\bibfnamefont {Henning}\ \bibnamefont
  {Rehbein}}, \bibinfo {author} {\bibfnamefont {Roman}\ \bibnamefont
  {Schnabel}}, \bibinfo {author} {\bibfnamefont {Karsten}\ \bibnamefont
  {Danzmann}}, \ and\ \bibinfo {author} {\bibfnamefont {Yanbei}\ \bibnamefont
  {Chen}},\ }\bibfield  {title} {\enquote {\bibinfo {title} {{Entanglement of
  Macroscopic Test Masses and the Standard Quantum Limit in Laser
  Interferometry}},}\ }\href {\doibase 10.1103/PhysRevLett.100.013601}
  {\bibfield  {journal} {\bibinfo  {journal} {Physical Review Letters}\
  }\textbf {\bibinfo {volume} {100}},\ \bibinfo {pages} {013601} (\bibinfo
  {year} {2008})}\BibitemShut {NoStop}%
\bibitem [{\citenamefont {Bar-Shalom}\ \emph {et~al.}(2001)\citenamefont
  {Bar-Shalom}, \citenamefont {Li},\ and\ \citenamefont
  {Kirubarajan}}]{bar-shalom_estimation_2001}%
  \BibitemOpen
  \bibfield  {author} {\bibinfo {author} {\bibfnamefont {Yaakov}\ \bibnamefont
  {Bar-Shalom}}, \bibinfo {author} {\bibfnamefont {X~Rong}\ \bibnamefont {Li}},
  \ and\ \bibinfo {author} {\bibfnamefont {Thiagalingam}\ \bibnamefont
  {Kirubarajan}},\ }\href@noop {} {\emph {\bibinfo {title} {{Estimation with
  Applications to Tracking and Navigation: Theory Algorithms and Software}}}}\
  (\bibinfo  {publisher} {John Wiley \& Sons},\ \bibinfo {year}
  {2001})\BibitemShut {NoStop}%
\bibitem [{\citenamefont {Heijden}\ \emph {et~al.}(2005)\citenamefont
  {Heijden}, \citenamefont {Duin}, \citenamefont {Ridder},\ and\ \citenamefont
  {Tax}}]{heijden_classification_2005}%
  \BibitemOpen
  \bibfield  {author} {\bibinfo {author} {\bibfnamefont {Ferdinand van~der}\
  \bibnamefont {Heijden}}, \bibinfo {author} {\bibfnamefont {Robert}\
  \bibnamefont {Duin}}, \bibinfo {author} {\bibfnamefont {Dick~de}\
  \bibnamefont {Ridder}}, \ and\ \bibinfo {author} {\bibfnamefont {David M~J}\
  \bibnamefont {Tax}},\ }\href@noop {} {\emph {\bibinfo {title}
  {Classification, Parameter Estimation and State Estimation: An Engineering
  Approach Using {MATLAB}}}}\ (\bibinfo  {publisher} {John Wiley \& Sons},\
  \bibinfo {year} {2005})\BibitemShut {NoStop}%
\bibitem [{\citenamefont {Stengel}(1994{\natexlab{b}})}]{stengel_optimal_1994}%
  \BibitemOpen
  \bibfield  {author} {\bibinfo {author} {\bibfnamefont {Robert~F}\
  \bibnamefont {Stengel}},\ }\href@noop {} {\emph {\bibinfo {title} {{Optimal
  Control and Estimation}}}},\ \bibinfo {edition} {reissue edition}\ ed.\
  (\bibinfo  {publisher} {Dover Publications},\ \bibinfo {address} {New York},\
  \bibinfo {year} {1994})\BibitemShut {NoStop}%
\bibitem [{\citenamefont {Clerk}\ \emph {et~al.}(2010)\citenamefont {Clerk},
  \citenamefont {Devoret}, \citenamefont {Girvin}, \citenamefont {Marquardt},\
  and\ \citenamefont {Schoelkopf}}]{clerk_introduction_2010}%
  \BibitemOpen
  \bibfield  {author} {\bibinfo {author} {\bibfnamefont {A~A}\ \bibnamefont
  {Clerk}}, \bibinfo {author} {\bibfnamefont {M~H}\ \bibnamefont {Devoret}},
  \bibinfo {author} {\bibfnamefont {S~M}\ \bibnamefont {Girvin}}, \bibinfo
  {author} {\bibfnamefont {Florian}\ \bibnamefont {Marquardt}}, \ and\ \bibinfo
  {author} {\bibfnamefont {R~J}\ \bibnamefont {Schoelkopf}},\ }\bibfield
  {title} {\enquote {\bibinfo {title} {{Introduction to quantum noise,
  measurement, and amplification}},}\ }\href {\doibase
  10.1103/RevModPhys.82.1155} {\bibfield  {journal} {\bibinfo  {journal}
  {Reviews of Modern Physics}\ }\textbf {\bibinfo {volume} {82}},\ \bibinfo
  {pages} {1155--1208} (\bibinfo {year} {2010})}\BibitemShut {NoStop}%
\bibitem [{\citenamefont {Brooks}\ \emph {et~al.}(2012)\citenamefont {Brooks},
  \citenamefont {Botter}, \citenamefont {Schreppler}, \citenamefont {Purdy},
  \citenamefont {Brahms},\ and\ \citenamefont
  {Stamper-Kurn}}]{brooks_non-classical_2012}%
  \BibitemOpen
  \bibfield  {author} {\bibinfo {author} {\bibfnamefont {Daniel W~C}\
  \bibnamefont {Brooks}}, \bibinfo {author} {\bibfnamefont {Thierry}\
  \bibnamefont {Botter}}, \bibinfo {author} {\bibfnamefont {Sydney}\
  \bibnamefont {Schreppler}}, \bibinfo {author} {\bibfnamefont {Thomas~P}\
  \bibnamefont {Purdy}}, \bibinfo {author} {\bibfnamefont {Nathan}\
  \bibnamefont {Brahms}}, \ and\ \bibinfo {author} {\bibfnamefont {Dan~M}\
  \bibnamefont {Stamper-Kurn}},\ }\bibfield  {title} {\enquote {\bibinfo
  {title} {{Non-classical light generated by quantum-noise-driven cavity
  optomechanics}},}\ }\href {\doibase 10.1038/nature11325} {\bibfield
  {journal} {\bibinfo  {journal} {Nature}\ }\textbf {\bibinfo {volume} {488}},\
  \bibinfo {pages} {476--480} (\bibinfo {year} {2012})}\BibitemShut {NoStop}%
\bibitem [{\citenamefont {Safavi-Naeini}\ \emph
  {et~al.}(2013{\natexlab{a}})\citenamefont {Safavi-Naeini}, \citenamefont
  {Gr\"{o}blacher}, \citenamefont {Hill}, \citenamefont {Chan}, \citenamefont
  {Aspelmeyer},\ and\ \citenamefont {Painter}}]{safavi-naeini_squeezed_2013}%
  \BibitemOpen
  \bibfield  {author} {\bibinfo {author} {\bibfnamefont {Amir~H}\ \bibnamefont
  {Safavi-Naeini}}, \bibinfo {author} {\bibfnamefont {Simon}\ \bibnamefont
  {Gr\"{o}blacher}}, \bibinfo {author} {\bibfnamefont {Jeff~T}\ \bibnamefont
  {Hill}}, \bibinfo {author} {\bibfnamefont {Jasper}\ \bibnamefont {Chan}},
  \bibinfo {author} {\bibfnamefont {Markus}\ \bibnamefont {Aspelmeyer}}, \ and\
  \bibinfo {author} {\bibfnamefont {Oskar}\ \bibnamefont {Painter}},\
  }\bibfield  {title} {\enquote {\bibinfo {title} {{Squeezed light from a
  silicon micromechanical resonator}},}\ }\href {\doibase 10.1038/nature12307}
  {\bibfield  {journal} {\bibinfo  {journal} {Nature}\ }\textbf {\bibinfo
  {volume} {500}},\ \bibinfo {pages} {185--189} (\bibinfo {year}
  {2013}{\natexlab{a}})}\BibitemShut {NoStop}%
\bibitem [{\citenamefont {Purdy}\ \emph
  {et~al.}(2013{\natexlab{b}})\citenamefont {Purdy}, \citenamefont {Yu},
  \citenamefont {Peterson}, \citenamefont {Kampel},\ and\ \citenamefont
  {Regal}}]{purdy_strong_2013}%
  \BibitemOpen
  \bibfield  {author} {\bibinfo {author} {\bibfnamefont {T~P}\ \bibnamefont
  {Purdy}}, \bibinfo {author} {\bibfnamefont {P-L}\ \bibnamefont {Yu}},
  \bibinfo {author} {\bibfnamefont {R~W}\ \bibnamefont {Peterson}}, \bibinfo
  {author} {\bibfnamefont {N~S}\ \bibnamefont {Kampel}}, \ and\ \bibinfo
  {author} {\bibfnamefont {C~A}\ \bibnamefont {Regal}},\ }\bibfield  {title}
  {\enquote {\bibinfo {title} {{Strong optomechanical squeezing of light}},}\
  }\href {http://journals.aps.org/prx/abstract/10.1103/PhysRevX.3.031012}
  {\bibfield  {journal} {\bibinfo  {journal} {Physical Review X}\ }\textbf
  {\bibinfo {volume} {3}},\ \bibinfo {pages} {031012} (\bibinfo {year}
  {2013}{\natexlab{b}})}\BibitemShut {NoStop}%
\bibitem [{\citenamefont {Safavi-Naeini}\ \emph
  {et~al.}(2013{\natexlab{b}})\citenamefont {Safavi-Naeini}, \citenamefont
  {Chan}, \citenamefont {Hill}, \citenamefont {Gr\"{o}blacher}, \citenamefont
  {Miao}, \citenamefont {Chen}, \citenamefont {Aspelmeyer},\ and\ \citenamefont
  {Painter}}]{safavi-naeini_laser_2013}%
  \BibitemOpen
  \bibfield  {author} {\bibinfo {author} {\bibfnamefont {Amir~H}\ \bibnamefont
  {Safavi-Naeini}}, \bibinfo {author} {\bibfnamefont {Jasper}\ \bibnamefont
  {Chan}}, \bibinfo {author} {\bibfnamefont {Jeff~T}\ \bibnamefont {Hill}},
  \bibinfo {author} {\bibfnamefont {Simon}\ \bibnamefont {Gr\"{o}blacher}},
  \bibinfo {author} {\bibfnamefont {Haixing}\ \bibnamefont {Miao}}, \bibinfo
  {author} {\bibfnamefont {Yanbei}\ \bibnamefont {Chen}}, \bibinfo {author}
  {\bibfnamefont {Markus}\ \bibnamefont {Aspelmeyer}}, \ and\ \bibinfo {author}
  {\bibfnamefont {Oskar}\ \bibnamefont {Painter}},\ }\bibfield  {title}
  {\enquote {\bibinfo {title} {{Laser noise in cavity-optomechanical cooling
  and thermometry}},}\ }\href {\doibase 10.1088/1367-2630/15/3/035007}
  {\bibfield  {journal} {\bibinfo  {journal} {New Journal of Physics}\ }\textbf
  {\bibinfo {volume} {15}},\ \bibinfo {pages} {035007} (\bibinfo {year}
  {2013}{\natexlab{b}})}\BibitemShut {NoStop}%
\bibitem [{\citenamefont {Riedinger}(2013)}]{riedinger_2013}%
  \BibitemOpen
  \bibfield  {author} {\bibinfo {author} {\bibfnamefont {Ralf}\ \bibnamefont
  {Riedinger}},\ }\emph {\bibinfo {title} {{Optomechanical State Reconstruction
  and Optical Noise Reduction for Cavity Optomechanics Experiments}}},\
  \href@noop {} {\bibinfo {type} {Master thesis}},\ \bibinfo  {school}
  {Philipps University Marburg}, \bibinfo {address} {Germany} (\bibinfo {year}
  {2013})\BibitemShut {NoStop}%
\bibitem [{\citenamefont {Gough}\ and\ \citenamefont
  {James}(2009)}]{gough_series_2009}%
  \BibitemOpen
  \bibfield  {author} {\bibinfo {author} {\bibfnamefont {J}~\bibnamefont
  {Gough}}\ and\ \bibinfo {author} {\bibfnamefont {MR}~\bibnamefont {James}},\
  }\bibfield  {title} {\enquote {\bibinfo {title} {{The Series Product and Its
  Application to Quantum Feedforward and Feedback Networks}},}\ }\href
  {\doibase 10.1109/TAC.2009.2031205} {\bibfield  {journal} {\bibinfo
  {journal} {Automatic Control, {IEEE} Transactions on}\ }\textbf {\bibinfo
  {volume} {54}},\ \bibinfo {pages} {2530 --2544} (\bibinfo {year}
  {2009})}\BibitemShut {NoStop}%
\bibitem [{\citenamefont {Nurdin}\ \emph {et~al.}(2009)\citenamefont {Nurdin},
  \citenamefont {James},\ and\ \citenamefont {Doherty}}]{nurdin_network_2009}%
  \BibitemOpen
  \bibfield  {author} {\bibinfo {author} {\bibfnamefont {H}~\bibnamefont
  {Nurdin}}, \bibinfo {author} {\bibfnamefont {M}~\bibnamefont {James}}, \ and\
  \bibinfo {author} {\bibfnamefont {A}~\bibnamefont {Doherty}},\ }\bibfield
  {title} {\enquote {\bibinfo {title} {{Network Synthesis of Linear Dynamical
  Quantum Stochastic Systems}},}\ }\href {\doibase 10.1137/080728652}
  {\bibfield  {journal} {\bibinfo  {journal} {{SIAM} Journal on Control and
  Optimization}\ }\textbf {\bibinfo {volume} {48}},\ \bibinfo {pages}
  {2686--2718} (\bibinfo {year} {2009})}\BibitemShut {NoStop}%
\bibitem [{\citenamefont {Leonhardt}(1993)}]{leonhardt_quantum_1993}%
  \BibitemOpen
  \bibfield  {author} {\bibinfo {author} {\bibfnamefont {Ulf}\ \bibnamefont
  {Leonhardt}},\ }\bibfield  {title} {\enquote {\bibinfo {title} {{Quantum
  statistics of a lossless beam splitter: {SU}(2) symmetry in phase space}},}\
  }\href {\doibase 10.1103/PhysRevA.48.3265} {\bibfield  {journal} {\bibinfo
  {journal} {Physical Review A}\ }\textbf {\bibinfo {volume} {48}},\ \bibinfo
  {pages} {3265--3277} (\bibinfo {year} {1993})}\BibitemShut {NoStop}%
\end{thebibliography}%

\end{document}